\documentclass[onecolumn]{emulateapj}
\usepackage{amsmath}
\usepackage{lineno}

\slugcomment{submitted; \today}

\newcommand{\p}{^\prime}
\newcommand{\e}{\epsilon}

\newcommand{\g}{\gamma}

\newcommand{\ep}{\epsilon^\prime}

\newcommand{\psim}{\lower.5ex\hbox{$\; \buildrel \propto \over\sim \;$}}
\newcommand{\lbar}{\lower.0ex\hbox{$\; \buildrel{\lower0.0ex \hbox{-}} \over\lambda  \;$}}

\newcommand{\m}{\mathrm{m}}
\newcommand{\cm}{\mathrm{cm}}
\newcommand{\km}{\mathrm{km}}
\newcommand{\erg}{\mathrm{erg}}
\newcommand{\Watt}{\mathrm{W}}
\newcommand{\dyne}{\mathrm{dyn}}
\newcommand{\gram}{\mathrm{g}}
\newcommand{\eV}{\mathrm{eV}}

\newcommand{\GeV}{\mathrm{GeV}}
\newcommand{\TeV}{\mathrm{TeV}}
\newcommand{\s}{\mathrm{s}}

\newcommand{\Mpc}{\mathrm{Mpc}}

\newcommand{\Kelvin}{\mathrm{K}}

\newcommand{\barZ}{\bar{Z}}

\shorttitle{Modeling the EBL}
\shortauthors{Finke et al.}

\begin{document}

\title{Modeling the Extragalactic Background Light and the Cosmic Star Formation
History}

\author{Justin D.\ Finke$^1$, Marco Ajello$^2$, Alberto Dom\'inguez$^3$,
  Abhishek Desai$^4$, Dieter H.\ Hartmann$^2$, 
  Vaidehi S.\ Paliya$^5$, Alberto Saldana-Lopez$^6$}

\affil{
$^1$ U.S.\ Naval Research Laboratory, Code 7653, 4555 Overlook Ave.\ SW,
     Washington, DC,
     20375-5352, USA; \\
     justin.finke@nrl.navy.mil \\
$^2$ Department of Physics and Astronomy, 
     Clemson University, 
     Kinard Lab of Physics,
     Clemson, SC 29634-0978, USA \\
$^3$ IPARCOS and Department of EMFTEL, 
     Universidad Complutense de Madrid, E-28040 Madrid, Spain \\
$^4$ Department of Physics and Wisconsin IceCube Particle Astrophysics Center, 
     University of Wisconsin-Madison, 
     Madison, WI 53706, USA \\
$^5$ Inter-University Centre for Astronomy and Astrophysics (IUCAA), 
     SPPU Campus, 411007, Pune, India\\
$^6$ Department of Astronomy, 
     University of Geneva, 
     51 chemin Pegasi, 1290 Versoix, Switzerland \\
}

\begin{abstract}

We present an updated model for the extragalactic background light
(EBL) from stars and dust, over wavelengths $\approx 0.1\ \mu\m$ to
$1000\ \mu\m$.  This model uses accurate theoretical stellar spectra,
and tracks the evolution of star formation, stellar mass
density, metallicity, and interstellar dust extinction and emission in
the universe with redshift.  Dust emission components are treated
self-consistently, with stellar light absorbed by dust reradiated in
the infrared as three blackbody components.  We fit our model, with
free parameters associated with star formation rate and dust
extinction and emission, to a wide variety of data: luminosity
density, stellar mass density, and dust extinction data from
galaxy surveys; and $\g$-ray absorption optical depth data from
$\g$-ray telescopes.  Our results strongly constraint the star
formation rate density and dust photon escape fraction of the universe
out to redshift $z=10$, about 90\% of the history of the universe.  We
find our model result is, in some cases, below lower limits on the
$z=0$ EBL intensity, and below some low-$z$ $\g$-ray absorption
measurements.  

\end{abstract}


\section{Introduction}
\label{intro}

The Extragalactic Background Light (EBL) in the range
$0.1-1000$\ $\mu$m consists of the background light from all of the
stars that have existed in the observable universe.  Direct stellar
emission produces a component at $\approx 0.1-4$ $\mu$m (the cosmic
optical background or COB), and the absorption of starlight that is
reradiated in the infrared (IR) produces a component at $\approx
4-1000$\ $\mu$m (the cosmic IR background or CIB).  The EBL contains a
great deal of information about the star formation history of the
universe and dust emission; however, it is difficult to observe
directly.  The night sky is dominated by emission from the atmosphere,
the solar system \citep[zodiacal light emitted by interplanetary
    dust mostly within the orbit of Jupiter;
    e.g.,][]{wright01,rowan13,korngut22}, and the Milky Way 
  \citep[e.g.,][]{seon11,brandt12,chellew22}, all of which are
brighter than the EBL.  Telescopes above the atmosphere can measure
the EBL without the contaminating atmospheric foreground
\citep[e.g.,][]{hauser98,bernstein02,mattila03,bernstein07}, 
  although these can still suffer from contamination from stray light
  from the Eath, Moon, or Sun outside the field of view of the
  instrument \citep{caddy22}.  Spacecraft beyond the orbit of Jupiter
have made measurements of the EBL with minimal contamination from the
zodiacal light
\citep{toller83,edelstein00,matsuoka11,zemcov17,lauer21,lauer22}.
However, there is still the difficulty of contamination from the Milky
Way foreground.  Galaxy number counts can give lower limits on the EBL
\citep[e.g.,][]{madau00,marsden09,driver16,koushan21} but do not
  include unresolved sources.

It was realized in the 1960s that the EBL would have an effect on
observed $\g$-ray spectra of extragalactic sources
\citep{nikishov62,gould67_EBL,fazio70}.  The high-energy $\g$ rays are
absorbed, producing electron positron pairs.  EBL photons with
wavelength $\lambda_{\rm EBL}$ will absorb $\g$ rays above observed
energy
\begin{flalign}
\label{energyeqn}
E_\g = 0.22\ \left( \frac{\lambda_{\rm EBL}}{\mu{\rm m}}\right)\ \TeV
\end{flalign}
producing electron-positron pairs and absorbing the $\g$-ray photon
\citep[e.g.,][]{gould67_crosssec}.  The cross section for this process
is strongly peaked at an energy about a factor of 2 greater
than the energy in Equation (\ref{energyeqn}).  
In recent years, this
has led to a number of attempts to measure the EBL with extragalactic
$\g$-ray sources, primarily blazars, and a number of modeling efforts.  

Constraints on the EBL with $\g$ rays have been found with
ground-based imaging atmospheric telescopes (IACTs) such as MAGIC,
H.E.S.S., and VERITAS; and with the Large Area Telescope (LAT) on the
{\em Fermi Gamma-ray Space Telescope} in low Earth orbit.  The
different telescopes observe at different energies, and probe
different wavelengths and redshifts ($z$) of the EBL.  The IACTs are
sensitive to absorption of $\sim 0.1$--$10\ \TeV$\ $\g$ rays, which
allows them to constrain the nearby ($z\la 1$) EBL at $\lambda_{\rm
  EBL}\approx 0.5$---$50\ \mu$m [Equation (\ref{energyeqn})].  The LAT
observes $\g$-ray absorption in the $10$---$100\ \GeV$ range
($\g$-rays below $\approx 10$\ GeV are not absorbed) and so probes the
EBL at $z\ga1$ and $\lambda_{\rm EBL}\approx 0.5$---$0.05\ \mu$m.
IACT constraints on the EBL can be done with IACTs only
\citep[e.g.,][]{aharonian99, protheroe00, schroedter05_EBL,
  aharonian06,aharonian07_EBL, mazin07, finke09, orr11,
  abramowski13_hessEBL, biteau15, desai19,
  abey19_veritasEBL,biasuzzi19} or extrapolating the LAT spectrum into
the IACT bandpass and using this to constrain the intrinsic blazar
very-high energy (VHE) $\g$-ray spectra
\citep[e.g.,][]{georgan10,meyer12_EBL,acciari19}; or by performing
detailed multiwavelength modeling of the blazar $\g$-ray source to
constrain the intrinsic VHE spectra
\citep[e.g.,][]{mank10,dominguez13_cgrh}.  LAT constraints on the EBL
have been found from blazars and gamma-ray bursts (GRBs) by
\citet{abdo10_EBL,ackermann12,desai17}; and \citet{abdollahi18}.
Besides $\g$-ray absorption, the EBL can create $\g$-rays in the LAT
bandpass, by being Compton scattered by high-energy electrons in the
lobes of radio galaxies.  This could in principle be used to constrain
the EBL through LAT observations of radio galaxies
\citep{georganopoulos08}; however, efforts to do so have been hampered
by unexpected additional $\g$-ray emission components from the radio
lobes, possibly of hadronic origin
\citep{mckinley15,ackermann16_fornaxA}, the existence of which is
controversial \citep{persic19_others,persic19_fornaxA,persic20}.  A
review of $\g$-ray constraints on the EBL is given by \citet{dwek13}.

Recent EBL models begin by modeling the ultraviolet (UV) through IR
luminosity densities (the total luminosity per unit volume of the
universe) across cosmic time; then integrating over $z$ to get the EBL
intensity as a function of redshift; then integrating over the
redshift between a $\g$-ray source and us to compute the $\g\g$
absorption optical depth (see Sections \ref{EBLsection} and
\ref{grayabsorbsection} below for details).  Modeling the EBL and
determining the model $\g$-ray absorption then is a matter of
determining the luminosity density.  The luminosity density can be
determined by integrating a galaxy luminosity function, which is
determined from deep surveys.  Models which directly use the results
of galaxy survey data to construct the EBL include the models of
\citet{franceschini08, dominguez11, helgason12, stecker12, scully14,
  stecker16,franceschini17}; and \citet{saldana21}.  Other models use
the cosmic star formation rate density (SFRD), models of stellar
emission, interstellar dust extinction, and dust emission to determine
the luminosity density.  The SFRD can be determined from various
measurements of star formation, as in the models of 
  \citet{salamon98, kneiske02, kneiske04, razzaque09, finke10_EBL,
    kneiske10, khaire15, andrews18,khaire19}, and \citet{koushan21}.
The SFRD can also be determined from semi-analytic models of galaxy
and large-scale structure formation, as in the models of
\citet{primack05, primack08, gilmore09,gilmore12_model}; and
\citet{inoue13}.  Recent models either use directly the luminosity
function or luminosity density measurements from galaxy surveys, or
attempt to reproduce them.  This has led to some amount of convergence
in recent years, with the models generally giving similar results,
indicating an EBL intensity at $z=0$ that is very close to the lower
limits from galaxy counts.

The close relationship between SFRD and the model $\g$-ray absorption
has led to many authors exploring using $\g$-ray absorption to put
(model-dependent) constraints on the SFRD \citep{raue12,gong13}.  This
includes putting constraints on the first generation of stars in the
early universe at high redshift, the ``Population III'' stars
\citep{raue09,gilmore12_pop3, inoue14}.  The SFRD is in turn closely
tied with the stellar mass density of the universe
\citep[e.g.,][]{wilkins08,perez08,madau14}, so that constraints on
luminosity density, SFRD, stellar mass density, and the EBL are
all intertwined \citep{fardal07}.

\citet{abdollahi18} presented LAT $\g$-ray opacity results from 739
blazars and 1 GRB.  They did a Markov Chain Monte Carlo (MCMC) fit 
the opacity data with two independent models to measure the SFRD
between $z=0$ and $z\approx6$.  Their results were consistent with
each other and with other star formation measures, most notably those
from luminosity densities.  

Here we do a much more expanded version of the model fit of
\citet{abdollahi18}.  We do a global model fit to the EBL opacity data
from the LAT and IACTs, and a wide variety of luminosity density, 
  stellar mass density, and dust extinction data from galaxy surveys,
taken from the literature, in order to provide tight constraints on
the SFRD.  We also include lower limits on the local EBL intensity.
We have also improved our EBL model for the fit, which was previously
described by \citet{razzaque09} and \citet{finke10_EBL}.  In Section
\ref{model} we describe the updated model, while in Section \ref{data}
we describe the wide variety of data to which we fit our model.  Our
results\footnote{The luminosity densities, EBL intensities, and
    $\g$-ray absorption optical depths from Model A from this paper
    have been made publicly available on Zenodo \citep{finke22}. } are
described in Section \ref{results} and we conclude with a discussion
in Section \ref{discussion}.

\section{EBL Model}
\label{model}

Our model is an extension of the one described by \citet{razzaque09}
and \citet{finke10_EBL}.  It has been upgraded in a number of ways.
The new model is fully described below.  The primary differences
between this model and the previous one are:
\begin{itemize}
\item Use PEGASE.2 stellar models.
\item Metallicity evolution with redshift.
\item Dust extinction evolution with redshift.
\item Different cosmic star formation rate density parameterizations.
\end{itemize}
We do not include the contribution of active galactic nuclei
(AGN) to the EBL.  Previous work has shown that this could contribute
up to 10\% of the UV through IR EBL
\citep{dominguez11,andrews18,abdollahi18,khaire19}.  Sections
\ref{cosmosection} through \ref{lumdenssection} describe the
components that go into computing the luminosity density.  Once the
luminosity density is known, the EBL intensity (or equivalently,
energy density) and $\g$-ray absorption optical depth can be computed,
as described in Sections \ref{EBLsection} and \ref{grayabsorbsection},
respectively.  The computations in Sections \ref{EBLsection} and
\ref{grayabsorbsection} will be the same for any EBL model, so that
the real work is in specifying how the luminosity density is
calculated.  The model has free parameters associated with the SFRD
and dust extinction and emission, which are allowed to vary.  We
describe them and the fit in Section \ref{MCMCsection}.

\subsection{Cosmology}
\label{cosmosection}

We use a flat $\Lambda$CDM cosmology, where in most of our models we
fix the parameters $H_0=70\ \km\ \s^{-1}\ \Mpc^{-1}$, $\Omega_m=0.30$
(where $\Omega_\Lambda= 1-\Omega_m = 0.70$ due to the flatness
assumption).  This cosmology is the most common one used for
luminosity density measurements found in the literature, with which we
compare our model results, and thus convenient for our purposes.
These cosmological values are also close to the ones independently
measured with a variety of methods.  There is a small but
statistically significant tension between the value of $H_0$ found by
using measures in the ``late'' universe (such as Type Ia supernovae
and lensed quasars) and in using anisotropy of the cosmic microwave
background (CMB) measured by Planck and other experiments
\citep[e.g.,][]{riess19,wong20,riess21,riess22}.  We make no attempt
to resolve this tension here; however, we note that absorption of $\g$
rays by EBL photons can be used to constrain the expansion rate of the
universe \citep[e.g.,][]{salamon94, mannheim96, blanch05, barrau08,
  fairbairn13, dominguez13, biteau15, zeng19, dominguez19}.  In
several of our model fits, we allow $H_0$ and $\Omega_m$ to be free
parameters.

We will make frequent use of the cosmological function
\begin{flalign}
\frac{dt}{dz} = \frac{-1}{H_0 (1+z)\sqrt{\Omega_m(1+z)^3 +
    \Omega_\Lambda}}\ ,
\end{flalign}
which relates a cosmological time interval to a redshift interval.

\subsection{Initial Mass Function}
\label{IMFsection}

The ``classic'' initial mass function (IMF) is that of
\citet{salpeter55}, given by
\begin{flalign}
\label{SalpeterIMFeqn}
\xi(m) = \frac{dN}{dm} = N_0 m^{-2.35}\ 
\end{flalign}
where $m$ is the stellar mass in $M_\odot$ units.  Although still
widely used for convenience, this IMF is now disfavored by
observations, especially at the low mass end
\citep[e.g.,][]{kroupa01,chabrier03}.  We use it in one of our
calculations, primarily to explore the effect of different IMFs on our
results.

\citet[][hereafter BG03]{baldry03} fit a stellar population synthesis
model with an IMF that was allowed to vary during their fit to a
collection of luminosity density data.  We primarily use their best fit
IMF, given by
\begin{equation}
\label{IMFeqn}
\xi(m) = \frac{dN}{dm} = N_0
\times \left\{
\begin{array}{ll}
(m/m_c)^{-1.5} & m \le m_c \\
(m/m_c)^{-2.2} & m > m_c 
\end{array}
\right.
\end{equation}
where $m_c=0.5$.

In both the Salpeter and BG03 cases, the constant $N_0$ is determined
by normalizing the IMF to a total mass of $M_\odot$, i.e.,
\begin{flalign}
1 = \int_{m_{\min}}^{m_{\max}} dm\ m\ \xi(m)\ .
\end{flalign}
In a future publication we will explore models with IMF parameters
free to vary in the fit.  We use $m_{\min}=0.1$ and $m_{\max}=120$.

\subsection{Recycling Stellar Material}
\label{recycle_section}

To compute the evolution of the mean metallicity in the ISM and the
stellar mass density one must know a number of quantities.  One
is the mass of the stellar remnant ($m_r(m,Z)$) at the end of a star's
life, whether a white dwarf, neutron star, or black hole.  This
quantity depends on both the progenitor star's mass ($m$) and
metallicity of the gas from which it was born ($Z$).  One must also
know the yield of new metals returned to the ISM for a star of a given
mass and birth metallicity, $p(m,Z)$.  From these we can compute the
quantities below.

The fraction of stellar mass returned to the ISM is calculated from
\citep[e.g.,][]{madau14,vincenzo16}
\begin{flalign}
\label{returnfrac}
R(Z) = \int_{m_{\rm ret}}^{m_{\max}} dm\ [m - m_r(m,Z)]\ \xi(m)
\end{flalign}
where $\xi(m)$ is the IMF.  The yield of heavy elements (the mass of
new heavy elements produced) returned to the ISM as a fraction of total mass
that is not returned to the ISM is
\begin{flalign}
\label{ytot}
y_{tot}(Z) = \frac{1}{1-R} \int_{m_{\rm ret}}^{m_{\min}} dm\ m\ p(m,Z)\ \xi(m)\ .
\end{flalign}
Here $m_{\rm ret}$ is the stellar mass above which metals are returned
to the ISM.  Both $R(Z)$ and $y_{tot}(Z)$ are fractions returned to
the ISM ineach stellar generation.

The quantities $R(Z)$ and $y_{\rm tot}(Z)$ were previously calculated
by \citet{vincenzo16} using a Salpeter IMF using $m_{\rm ret}=1.0$ and
$m_r(m,Z)$ and $p(m,Z)$ as calculated and tabulated by
\citet{nomoto13}.  We use these same method to compute $R(Z)$
and $y_{tot}(Z)$ for the BG03 IMF used here.  Our results can be found
in Table \ref{metaltable}.  For values of $Z$ not in Table
\ref{metaltable}, we did a linear interpolation between these values
to determine $R(Z)$ and $y_{tot}(Z)$.  These quantities are only
weakly dependent on $Z$ so there should be little uncertainty from
this interpolation.

\begin{deluxetable}{ccc}
\tablecaption{Stellar mass return fraction ($R(Z)$; Equation
  [\ref{returnfrac}]) and total stellar metal yield ($y_{tot}(Z)$;
  Equation [\ref{ytot}]) for the BG03 IMF.
}
\tablewidth{0pt}
\tablehead{
\colhead{$Z$} & 
\colhead{$R(Z)$} &
\colhead{$y_{tot}(Z)$}
}
\startdata
$0.0$ & $0.452$ & $0.108$ \\
$0.001$ & $0.498$ & $0.0653$ \\
$0.004$ & $0.507$ & $0.0631$ \\
$0.008$ & $0.511$ & $0.0616$ \\
$0.02$ & $0.516$ & $0.0588$ \\
$0.05$ & $0.515$ & $0.0686$ 
\enddata
\label{metaltable}
\vspace{3mm}
\end{deluxetable}

\subsection{Metallicity and stellar Mass Density Evolution}

To determine the ISM mean metallicity $\barZ(z)$ and comoving stellar
mass density $\rho(z)$ of the universe, we use a one-zone, closed box,
instantaneous recycling model \citep{tinsley80}.  This model assumes
that stars with masses $m<m_{\rm ret}$ will live forever and never
return matter to the ISM; and stars with masses $m>m_{\rm ret}$ will
instantly return mass and newly-formed metals to the ISM
\citep[e.g.,][]{madau14,vincenzo16}.  This model is a good
approximation for metals produced primarily by massive, short-lived
stars (such as oxygen) but less good for metals produced primarily by
low-mass, long-lived stars.  Since oxygen is the most abundant heavy
element by mass, this model should be a reasonable approximation
for mean metallicity \citep{vincenzo16}.

The quantities $\barZ(z)$ and $\rho(z)$ are found by solving the
equations \citep[e.g.,][]{madau14}
\begin{flalign}
\frac{d\barZ}{dz} & = \frac{y_{tot}(\barZ) [1 - R(\barZ)] }{\rho_b - \rho(z)}
\psi(z) \left| \frac{dt}{dz}\right| \ ,
\nonumber \\
\frac{d\rho}{dz} & = (1 - R(\barZ)) \psi(z) \left| \frac{dt}{dz}\right| \ ,
\label{evoequation}
\end{flalign}
respectively, where $R(\barZ)$ and $y_{tot}$ are described in Section
\ref{recycle_section}, $\psi(z)$ is the comoving SFRD and $\rho_b$ is
the total comoving mass density of baryons (in stars and gas) in the
universe.  It can be found from $\rho_b = \rho_c\Omega_b$ where
\begin{flalign}
\rho_c = \frac{3H_0}{8\pi G}\ ,
\end{flalign}
and $G=6.673\times10^{-8}\ \dyne\ \cm^{2}\ \gram^{-2} =
6.673\times10^{-11}\ {\rm N}\ \m^{2}\ {\rm kg}^{-2}$\ is the
gravitational constant.  The parameter $\Omega_b$ can be determined
from anisotropy in the CMB.  We use $\Omega_b=0.045$ consistent with
results from {\em WMAP} \citep{hinshaw13}.

The coupled ordinary differential equations (\ref{evoequation}) were solved with
a fourth order Runge-Kutta numerical scheme \citep[e.g.,][]{press92} with
the initial conditions $\barZ(z=z_{\max})=0$ and $\rho(z=z_{\max})=0$.

\subsection{Simple Stellar Population Spectra}

\begin{figure}
\vspace{2.2mm} 
\epsscale{1.0} 
\plotone{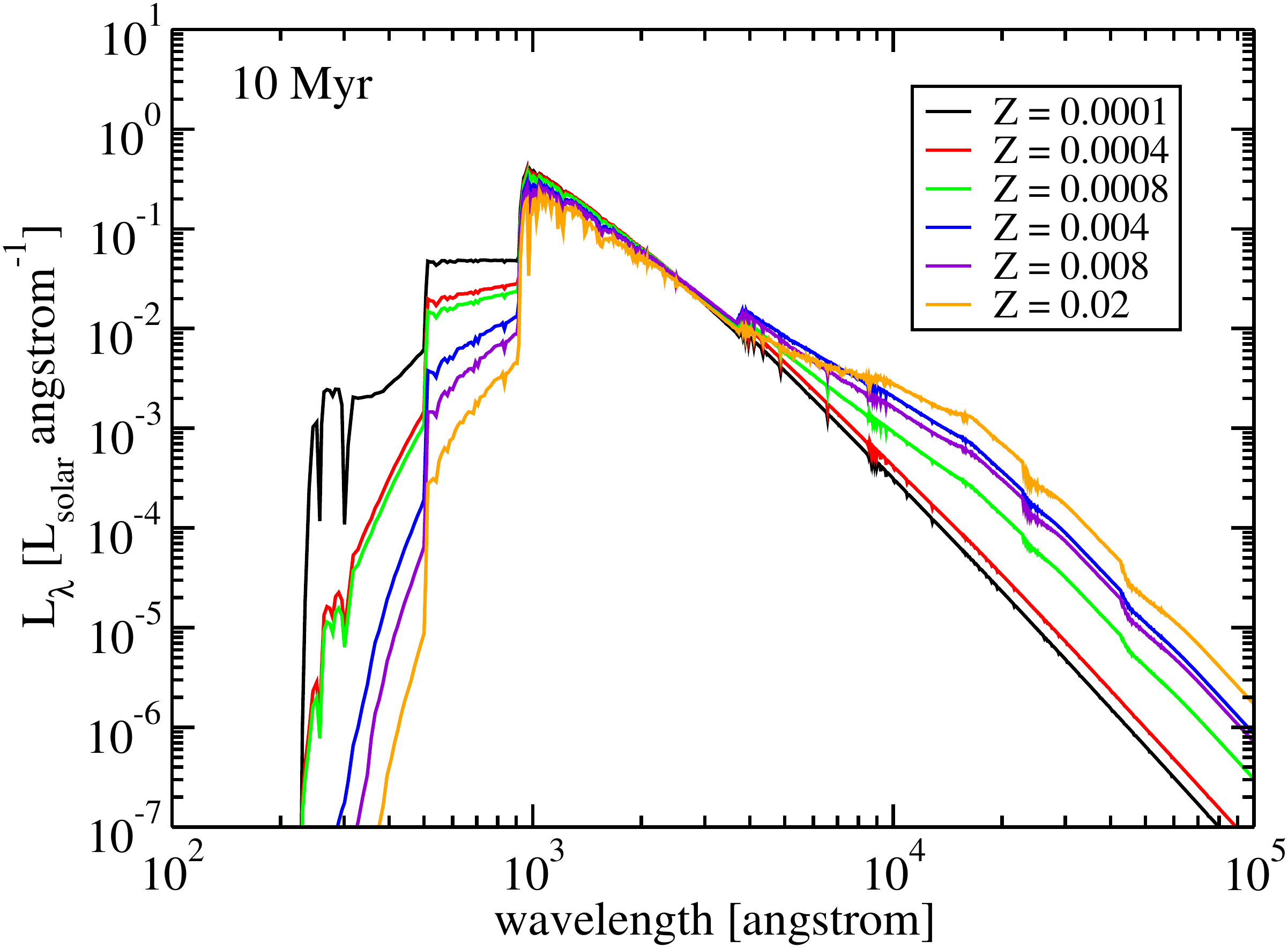}
\caption{SSPS for $t=10$\ Myr for different metallicities.  IR
  emission increases and UV emission decreases for increasing
  metallicity ($Z$).}
\label{ssp_compare}
\vspace{2.2mm}
\end{figure}

We use the PEGASE.2 code\footnote{\url{http://www2.iap.fr/pegase/}}
\citep{fioc97} to generate simple stellar population spectra (SSPS) of
a population of stars with masses between $m_{\min}$ and $m_{\max}$ as
a function of metallicity ($Z$) and age ($t$).  This assumes all stars
are born instantly at the same time at $t=0$, and evolve passively
with no further star formation.  The SSPS $L_\lambda(t, Z)$ are computed
in units $\erg\ \s^{-1}\ \AA^{-1}$ and are normalized to $1\ M_\odot$.
The PEGASE.2 code allows the user to generate SSPS for user defined
IMFs.  As discussed in Section \ref{IMFsection}, we use the BG03 IMF,
Equation (\ref{IMFeqn}), and the Salpeter IMF, Equation
(\ref{SalpeterIMFeqn}).  The primary effect of metallicity on the SSPS
is that at higher metallicity, there is more IR emission and less UV
emission (Figure \ref{ssp_compare}).

The PEGASE.2 code does not produce accurate
SSPS for high mass Population III (low $Z$) stars.  Therefore, for
SSPS at $Z=0$, we follow \citet{gilmore12_pop3} and use
the results from \citet{schaerer02} for high mass stars.  For $Z=0$
stars with $m<5$, we use the PEGASE.2 results; for $m>5$, we assume the
stars are blackbodies with temperatures and luminosities given by
Table 3 of \citet{schaerer02} for lifetimes given by Table 4 of
\citet{schaerer02}.  

Once SSPS are produced for a grid of stellar population ages and
metallicities, we interpolate to find the SSPS $L_\lambda(t, Z)$ for
any $t$ and $Z$.

\subsection{Dust Extinction}

A fraction $f_{\ esc,d}(\lambda, z)$ of photons will escape absorption
by dust.  \citet{driver08} calculated the average of this fraction at
$z\approx 0$ from fitting a dust model to 10,000 nearby galaxies.  The
resulting fraction at $z\approx 0$ fit by \citet{razzaque09},
resulting in
\begin{flalign}
f_{\rm esc,d}(\lambda, z = 0) =
 \left\{ 
\begin{array}{ll}
0.688 + 0.556\log_{10}\lambda; & \lambda \le 0.167 \\ 
0.151 - 0.136\log_{10}\lambda; & 0.167 < \lambda \le 0.218 \\
1.000 + 1.148\log_{10}\lambda; & 0.22 < \lambda \le 0.422 \\
0.728 + 0.422\log_{10}\lambda; & 0.422 < \lambda
\end{array}
\right.\ ,
\end{flalign}
where $\lambda$ is wavelength in $\mu$m.  The dust extinction evolves
with redshift; \citet{abdollahi18} fit a collection of measurements
with a curve as a function of $z$.  We assume the normalization of
$f_{\rm f esc,d}(\lambda,z)$ evolves with $z$ following this curve, so
that
\begin{flalign}
f_{\rm esc,d}(\lambda,z) = \frac{f_{\rm esc,d}(\lambda,z=0)}
{f_{\rm esc,d}(\lambda=0.15\ \mu\m,z=0)}
 \times10^{-0.4A(z)}
\end{flalign}
where
\begin{flalign}
\label{UVabscoeff}
A(z) = m_d \frac{ (1+z)^{n_d} }{1 + [(1+z)/p_d]^{q_d}}\ ,
\end{flalign}
with the \citet{abdollahi18} fit results $m_d=1.49$, $n_d=0.64$,
$p_d=3.4$, $q_d=3.54$.  A similar fit was done by \citet{puchwein19}
using a slightly different functional form.

We also use a dust model where
\begin{flalign}
f_{\rm esc,d}(\lambda, z = 0) =
 \left\{ 
\begin{array}{ll}
f_{\rm esc,2} + (f_{\rm esc,2}-f_{\rm esc,1})/
(\log_{10}\lambda_2-\log_{10}\lambda_1)\log_{10}\lambda; 
& \lambda \le \lambda_2 \\ 
f_{\rm esc,3} + (f_{\rm esc,3}-f_{\rm esc,2})/
(\log_{10}\lambda_3-\log_{10}\lambda_2)\log_{10}\lambda; 
& \lambda_2 < \lambda \le \lambda_3 \\ 
f_{\rm esc,4} + (f_{\rm esc,4}-f_{\rm esc,3})/
(\log_{10}\lambda_4-\log_{10}\lambda_3)\log_{10}\lambda; 
& \lambda_3 < \lambda \le \lambda_4 \\ 
f_{\rm esc,5} + (f_{\rm esc,5}-f_{\rm esc,4})/
(\log_{10}\lambda_5-\log_{10}\lambda_4)\log_{10}\lambda; 
& \lambda_4 < \lambda  \\ 
\end{array}
\right.
\end{flalign}
where $\lambda_1=0.15\ \mu$m, $\lambda_2=0.167\ \mu$m,
$\lambda_3=0.218\ \mu$m, $\lambda_4=0.422\ \mu$m,
$\lambda_5=2.0\ \mu$m.  Here $\{ f_{\rm esc,k}, k=1..5 \}$ 
are free parameters.

\subsection{Star Formation}

We use two parameterizations for the comoving SFRD as a function of
redshift.  We primarily use the formulation from \citet{madau14},
\begin{flalign}
\label{SFRmadau}
\psi_{\rm MD}(z) = a_s \frac{ (1+z)^{b_s} }{1 + [(1+z)/c_s]^{d_s}}
\end{flalign}
where $a_s$, $b_s$, $c_s$, and $d_s$ are free parameters.  We also
use a version similar to the piecewise function from \citet{hopkins06}, 
\begin{flalign}
\psi_{\rm piece}(z)  & =  10^{a_s}
 \left\{ 
\begin{array}{ll}
\zeta^{b_s} & z<z_1 \\
\zeta_1^{b_s-c_s}\zeta^{c_s} & z_1 \le z < z_2 \\
\zeta_1^{b_s-c_s}\zeta_2^{c_s-d_s}\zeta^{d_s} & z_2 \le z < z_3 \\
\zeta_1^{b_s-c_s}\zeta_2^{c_s-d_s}\zeta_3^{d_s-e_s}\zeta^{e_s} & z_3 \le z < z_4 \\
\zeta_1^{b_s-c_s}\zeta_2^{c_s-d_s}\zeta_3^{d_s-e_s}\zeta_4^{e_s-f_s}\zeta^{f_s} & z_3 \le z  \\
\end{array}
\right.
\label{piecewiseSFReqn}
\end{flalign}
where $a_s$, $b_s$, $c_s$, $d_s$, $e_s$, and $f_s$ are free
parameters, and we use the notation $\zeta=(1+z)$ and $\{
\zeta_k=(1+z_k), k=1..4\}$.  We fix $z_1=1.0$, $z_2=2.0$, $z_3=3.0$,
$z_4=4.0$.

\subsection{Luminosity Density}
\label{lumdenssection}

The comoving stellar luminosity density (i.e., luminosity per unit comoving volume) 
is given as a function of comoving dimensionless energy $\e=hc/(\lambda m_e c^2)$ by
\begin{flalign}
\e j^{\rm stars}(\e; z) & = m_ec^2\e^2 \frac{dN}{dtd\e dV}   
\nonumber \\ & =
f_{\rm esc,H}(\e) f_{\rm esc,d}(\lambda,z)\ 
\nonumber \\ & \times
\int_z^{z_{\max}} dz_1\ L_\star(t_\star(z,z_1), Z(z_1))\ \psi(z_1) \left| \frac{dt}{dz_1}\right|\ ,
\end{flalign}
where $L_\star(t,Z) = \lambda\times L_\lambda(t,Z)$.  We assume all photons at
energies greater than 13.6 eV are absorbed by \ion{H}{1} gas, so that
\begin{equation}
f_{\rm esc,H}(\e) = 
 \left\{
\begin{array}{ll}
1 & m_ec^2\e < 13.6\ \eV \\
0 & m_ec^2\e \ge 13.6\ \eV
\end{array}
\right.\ .
\end{equation}
The age of the stellar population $t_\star(z,z_1)$ can be found by
computing the integral
\begin{flalign}
t_\star(z,z_1) = \int_z^{z_1} dz\p\ \left| \frac{dt}{dz\p} \right| \ ,
\end{flalign}
which has an analytic solution \citep{razzaque09}.

We assume that the total energy absorbed by dust is re-emitted in the
infrared in three blackbody dust components.  These three components
are a large grain component with temperature $T_1$ (left as a
free parameter), a small
grain component with temperature $T_2=70$\ K, and a component
representing polycyclic aromatic hydrocarbons as a blackbody with
temperature $T_3=450$\ K.  The comoving dust luminosity density is
then
\begin{flalign}
\e j^{\rm dust}(\e; z) & = \frac{15}{\pi^4} \int d\e 
\left[ \frac{1}{f_{\rm esc,d}(\lambda,z)} - 1\right] j^{\rm stars}(\e; z)
\nonumber \\ & \times
\sum_{n=1}^3 \frac{f_n}{\Theta_n^4} \frac{\e^4}{\exp[\e/\Theta_n] - 1}
\end{flalign}
where $\Theta_n=k_B T_n/(m_ec^2)$ is the dimensionless temperature of
a particular component and $f_n<1$ is the fraction of the emission
absorbed by dust that is reradiated in a particular dust component.
We use $f_1$ and $f_2$ as free parameters, with $f_3$ constrained by
$f_1 + f_2 + f_3 = 1$.  The temperatures of each component are fixed
to the values given above.

Once the stellar and dust luminosity densities are calculated, 
the total luminosity density is simply
\begin{flalign}
\e j^{\rm tot}(\e; z) =  \e j^{\rm stars}(\e; z) + \e j^{\rm dust}(\e; z)\ .
\end{flalign}

\subsection{EBL}
\label{EBLsection}

The proper frame energy density as a function of the proper
frame dimensionless energy $\e_p$ is
\begin{flalign}
\e_p u_{\rm EBL,p}(\e_p; z) = (1+z)^4 \int_z^{z_{\max}} dz_1 
\frac{\ep j^{\rm tot}(\ep; z_1)}{(1+z_1)} \left| \frac{dt}{dz}\right|\ ,
\end{flalign}
where
\begin{flalign}
\ep = \frac{1+z_1}{1+z}\e_p\ .
\end{flalign}
EBL intensity is usually measured in units nW m$^{-2}$\ sr$^{-1}$; the
energy density can be converted to intensity via
\begin{flalign}
\e_p I_{\rm EBL}(\e_p; z) = \frac{c}{4\pi} \e_p u_{\rm EBL,p}(\e_p; z)\ .
\end{flalign}

\subsection{Gamma-ray Absorption}
\label{grayabsorbsection}

The absorption optical depth
\begin{flalign}
\label{taugg}
\tau_{\g\g}(\e_1, z) & = \frac{c\pi r_e^2}{\e_1^2 m_ec^2}\ 
\int^z_0 \frac{dz^\prime}{(1+z^\prime)^2}\ 
\left| \frac{dt_*}{dz^\prime}\right|\ 
\nonumber \\ & \times
\int^\infty_{\frac{1}{\e_1(1+z^\prime)}} d\e_{p} 
\frac{ \e_{p}u_{EBL,p}(\e_{p};z^\prime)}
{\e_p^{4}}\ 
\bar{\phi}(\e_{p}\e_1(1+z^\prime))\ ,
\end{flalign}
where
\begin{eqnarray}
\bar{\phi}(s_0) = \frac{1+\beta_0^2}{1-\beta_0^2}\ln w_0 - 
\beta_0^2\ln w_0 - \frac{4\beta_0}{1 - \beta_0^2}
\\ \nonumber 
+ 2\beta_0 + 4\ln w_0 \ln(1+w_0) - 4L(w_0)\ ,
\end{eqnarray}
$\beta_0^2 = 1 - 1/s_0$, $w_0=(1+\beta_0)/(1-\beta_0)$, and 
\begin{eqnarray}
L(w_0) = \int^{w_0}_1 dw\ w^{-1}\ln(1+w)\ 
\end{eqnarray}
\citep{gould67_crosssec,brown73}.  

\section{Data}
\label{data}

We fit a variety of data with our model using an MCMC algorithm.  The
data include $\g$-ray absorption data from blazars, and luminosity
density, stellar mass density, EBL intensity, and dust
extinction data from galaxy surveys.  We briefly describe these data
in Sections \ref{graydatasection}, \ref{LDdatasection},
\ref{MDdatasection}, \ref{EBLdatasection}, and
\ref{dustextinctiondatasection}, respectively, and the MCMC technique
in Section \ref{MCMCsection}.

\subsection{Gamma-ray Absorption Data}
\label{graydatasection}

\citet{ackermann12} developed a method to determine the EBL absorption
optical depth from $\g$-ray observations of a large sample of blazars.
This involved a joint likelihood fit to a large statistical sample of
blazars, using the LAT spectrum in the $1.0$---$10\ \GeV$ (where the
$\g$-ray absorption should be negligible) extrapolated to $>10\ \GeV$
as the intrinsic spectrum of each source, and allowing the
normalization of the absorption optical depth to be an additional free
parameter for all sources within a redshift and energy bin.  They
applied their method to 150 blazars in 3 redshift bins, using
$\approx4$ years of data.  \citet{abdollahi18} applied this technique
to a much larger sample and much more data: 9 years of LAT data from
739 blazars, now divided into 12 redshift bins.  They also included a
constraint from the LAT detection of GRB 080916C \citep[see
  also][]{desai17}.  \citet{desai19} applied this method to measure
the absorption optical depth using 106 IACT VHE $\g$-ray spectra from
38 blazars from the sample compiled by \citet{biteau15}.  They divided
their sample into two redshift bins.  We include in our fit here the
absorption optical depth measured with the 739 blazars and 1 GRB
measured by \citet{abdollahi18}, and measured with the 106 VHE spectra
by \citet{desai19}.  The reported errors for all of the $\g$-ray
absorption data include both statistical and systematic uncertainties.

Generally, $\g$-ray telescopes are only able to measure values of the
EBL absorption optical depth in the range $10^{-2}\la \tau_{\g\g} \la
5$.  This is because, for low values of $\tau_{\g\g}$, e.g.,
$\exp(-0.01)\approx0.99$, absorption will be negligible, and there
will be nothing to measure.  At high values of $\tau_{\g\g}$, e.g.,
$\exp(-5)\approx0.007$, almost all of the $\g$-rays will be absorbed,
and there will be no $\g$-ray signal to measure.  Outside of this
range, however, it is possible to put upper limits on $\tau_{\g\g}$
at lower optical depths, and lower limits on $\tau_{\g\g}$ at higher
optical depths.  

One should also note that Lorentz invariance violation could modify
the $\g\g$ pair production threshold, reducing the absorption optical
depth to $\g$-rays at $\ga 10\ \TeV$ \citep[e.g.,][]{kifune99}.  In
this cases, $\g$-rays at these energies could be potentially
detectable with the upcoming Cherenkov Telescope Array
\citep{abdalla18,abdalla21}.  If axion-like particles exist,
$\g$-ray photons could convert to them and in principle avoid $\g\g$
pair production \citep{deang07,sanchez09}.  However, there is no
evidence that this is occuring \citep{buehler20}.

\subsection{Luminosity Density Data}
\label{LDdatasection}

\begin{figure}
\vspace{2.2mm} 
\epsscale{1.0} 
\plotone{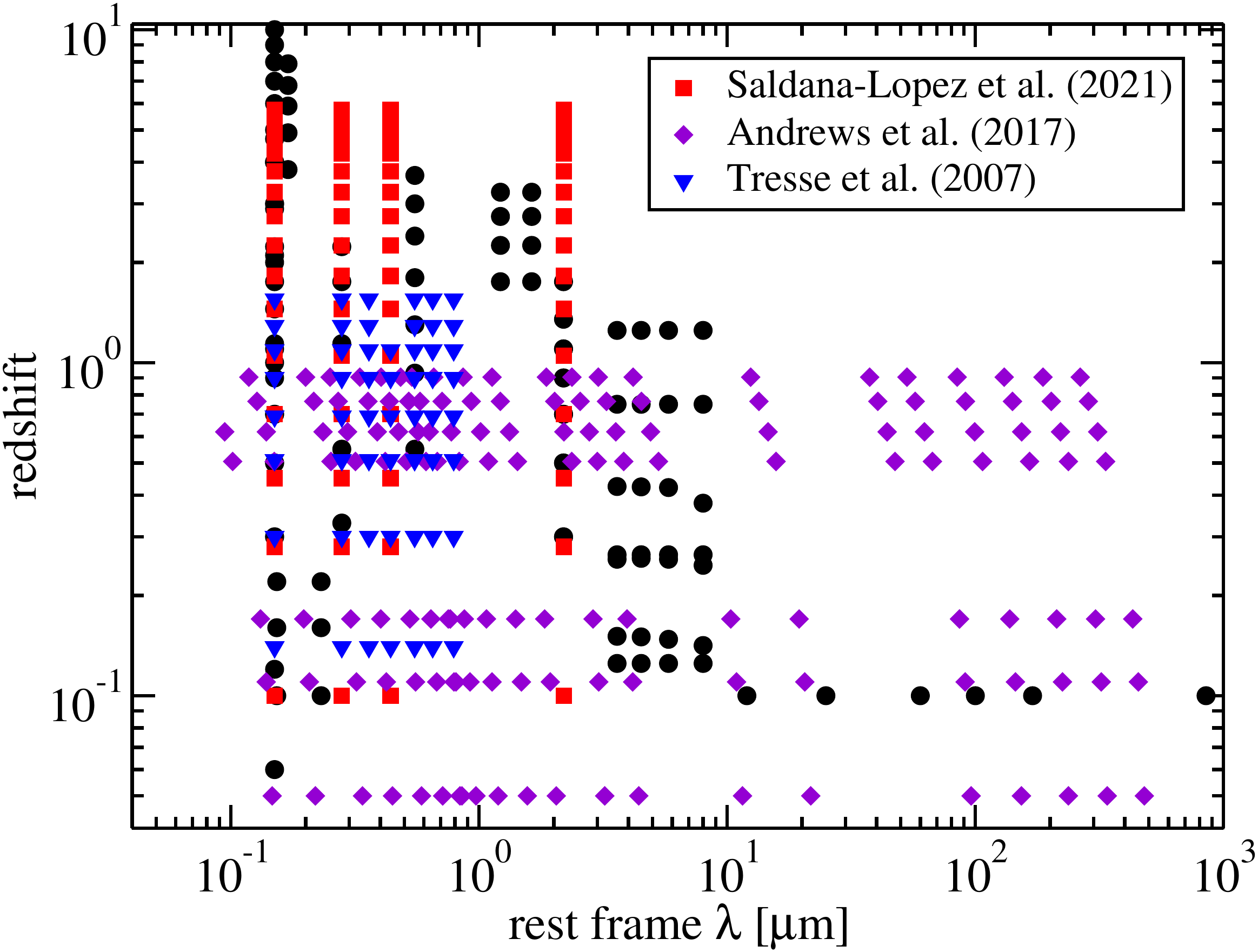}
\caption{A plot indicating the rest-frame wavelength and redshift of
  the luminsity density data from the literature that we use.  Data
  from \citet{tresse07,andrews17}; and \citet{saldana21} are
  represented by symbols as shown in the legend; all other sources are
  represented by circles. }
\label{lambda_z_cover_fig}
\vspace{2.2mm}
\end{figure}

We have searched through the literature to compile a large amount of
luminosity density measurements from galaxy surveys to fit with our
EBL model.  We particularly found useful the compilations by
\citet{helgason12,stecker12,scully14,madau14}; and \citet{stecker16}.
Our luminosity data compiled from the literature can be found in Table
\ref{lumdenstable}.  The redshift and wavelength coverage can be seen
in Figure \ref{lambda_z_cover_fig}.  The majority of the data comes
from three sources: \citet{tresse07,andrews17}; and \citet{saldana21}.
Coverage is fairly complete at $z\la 1$, mainly due to the work of
\citet{andrews17} using data from the Galaxy and Mass Assembly (GAMA)
and Cosmic Origins Survey (COSMOS) projects.  They included data from
a wide variety of instruments across the electromagnetic spectrum.
There is less complete coverage at longer wavelengths moving to higher
$z$.  \citet{saldana21} made use of the CANDELS programs to get fairly
complete wavelength coverage of the luminosity density up to $z=6$.
At $z\ga6$, only UV luminosity density data is available.  These
high-$z$ UV data are due to deep exposures with the WFC3 instrument on
the {\em Hubble Space Telescope (HST)}
\citep{finkelstein15,mcleod16,bouwens16}.  However, see
\citet{oesch18}.  Much of the luminosity density data we use here was
published after our last model \citep{finke10_EBL}, giving the work
here a different view of the luminosity density, particularly at
high redshift.  Occasionally we needed to convert luminosity density
from $L_\odot\ \Mpc^{-3}$ to $\Watt\ \Mpc^{-3}$, which we did using
$L_\odot = 3.826\times10^{26}\ \Watt$.

When necessary we converted the luminosity density measurements to
a cosmology with $H_0=70\ \km\ \s^{-1}\ \Mpc^{-1}$ and
$\Omega_m=0.30$ assumed in most of our models (Section
\ref{cosmosection}).  In several of our model fits (Models A.c and
D.c), we allow these cosmological parameters to be free parameters.
In these cases, the model luminosity density is converted from the
model cosmology to the ``standard''
$(H_0,\Omega_m)=(70\ \km\ \s^{-1}\ \Mpc^{-1}, 0.30)$ cosmology used in
the measurements with
\begin{flalign}
j(\e,z)_{\rm std} = j(\e,z)_{\rm model} \times 
\frac{ \left| dt/dz\right|_{\rm model} }{ \left| dt/dz\right|_{\rm std} }\ .
\end{flalign}

\begin{deluxetable}{cccr}
\tablecaption{Luminosity density data used in fits.  
}
\tablewidth{0pt}
\tablehead{
\colhead{$z$} & 
\colhead{$\lambda\ [\mu\m]$} &
\colhead{$\e j(\e;z)\ [\Watt\ \Mpc^{-3}]$} &
\colhead{Reference}
}
\startdata
     0.0   &    0.1535	&   $0.84^{+ 0.07}_{-0.07}\times10^{34}$ & $^a$ \\
     0.0   &    0.2301	&   $0.84^{+ 0.07}_{-0.07}\times10^{34}$ & $^a$ \\
     0.0   &    0.3557	&   $1.47^{+ 0.07}_{-0.07}\times10^{34}$ & $^a$ \\
     0.0   &    0.4702	&   $3.29^{+ 0.07}_{-0.07}\times10^{34}$ & $^a$ \\
     0.0   &    0.6175	&   $4.41^{+ 0.14}_{-0.14}\times10^{34}$ & $^a$ \\
     0.0   &    0.7491	&   $4.62^{+ 0.14}_{-0.14}\times10^{34}$ & $^a$ \\
     0.0   &    0.8946	&   $4.69^{+ 0.14}_{-0.14}\times10^{34}$ & $^a$ \\
     0.0   &    1.0305	&   $4.20^{+ 0.21}_{-0.21}\times10^{34}$ & $^a$ \\
     0.0   &    1.2354	&   $3.78^{+ 0.21}_{-0.21}\times10^{34}$ & $^a$ \\
     0.0   &    1.6458	&   $3.78^{+ 0.21}_{-0.21}\times10^{34}$ & $^a$ 
\enddata

\tablecomments{$^a$\citet{driver12}. $^b$\citet{andrews17}. 
$^c$\citet{tresse07}. $^d$\citet{arnouts07}. $^e$\citet{beare19}. 
$^f$\citet{budavari05}. $^g$\citet{cucciati12}. $^h$\citet{dahlen07}. 
$^i$\citet{marchesini12}. $^j$\citet{schimin05}. $^k$\citet{stefanon13}. 
$^l$\citet{bouwens16}.
$^m$\citet{finkelstein15}. $^n$\citet{mcleod16}. $^o$\citet{dai09}.
$^p$\citet{babbedge06}. 
$^q$\citet{takeuchi06}. $^r$\citet{saldana21}. $^s$\citet{yoshida06}.
\\
\\ 
This table is published in its entirety in the
  machine-readable format.  A portion is shown here for guidance
  regarding its form and content.}

\label{lumdenstable}
\end{deluxetable}

\subsection{Stellar Mass Density Data}
\label{MDdatasection}

\citet{madau14} have compiled a large number of cosmological 
  stellar mass density measurements.  Stellar mass density
measurements are model-dependent, particularly on the IMF.
\citet{madau14} have scaled their mass density calculation to a
Salpeter IMF.

We use stellar mass density data from the compilation by
\citet{madau14}.  For models that use the BG03 IMF, we adjusted the
stellar mass density as follows:
\begin{flalign}
\rho_{\rm BG03} = \rho_{\rm Sal} \times \frac{1 - R_{\rm BG03}}{1 - R_{\rm Sal}}
\end{flalign}
where $\rho_{\rm BG03}$ and $\rho_{\rm Sal}$ are the mass density data
with the BG03 and Salpeter IMFs, respectively, and $R_{\rm BG03}$ and
$R_{\rm Sal}$ are the return fractions (Equation \ref{returnfrac}) for
the BG03 and Salpeter IMFs, respectively.  We use $R_{\rm BG03}=0.49$
and $R_{\rm Sal}=0.3$ for all $z$ and $\bar Z$, although the actual
values are metallicity-dependent, these are roughly the midpoint of
the different values (Table \ref{metaltable}).  To take into account
the errors from assuming all stars have the same $R$, we increased the
error on all $\rho_{\rm BG03}$ by 10\%, added in quadrature; that is,
the resulting error in $\rho_{\rm BG03}$ is
\begin{flalign}
\sigma_{\rm BG03} = \sqrt{ \sigma_{\rm Sal}^2 + (0.1\rho_{\rm BG03})^2 }
\end{flalign}
where $\sigma_{\rm Sal}$ is the error on $\rho_{\rm Sal}$ reported by
\citet{madau14}.

Similar to the luminosity density, when the cosmological parameters
are allowed to vary in the model, we convert the model stellar
mass density to the standard cosmology used in the stellar mass
density measurements with
\begin{flalign}
\rho(z)_{\rm std} = \rho(z)_{\rm model} \times 
\frac{ \left| dt/dz\right|_{\rm model} }{ \left| dt/dz\right|_{\rm std} }\ .
\end{flalign}

\subsection{EBL intensity constraints at z=0}
\label{EBLdatasection}

We include in our model fit the $z=0$ EBL intensity integrated galaxy
light lower limits from \citet{driver16}.  This integrated galaxy
light represents the resolved component of the EBL.  The lower limits
span a wavelength range $0.15\ \mu\m < \lambda < 500\ \mu\m$ and were
derived from galaxy number count data from a variety of ground- and
space-based sources.

\subsection{UV Dust Extinction Data}
\label{dustextinctiondatasection}

We use the far ultraviolet (FUV; $0.15\ \mu\m$) dust extinction as
measured by \citet{burgarella13} and \citet{andrews17}.
\citet{burgarella13} derive the extinction from the ratio of the FUV
to integrated infrared luminosity functions.  \citet{andrews17}
determined these FUV extinction values by fitting a dust model to a
variety of multiwavelength data.

\subsection{MCMC fit}
\label{MCMCsection}

We use the open-source python-implemented MCMC code {\tt emcee}
\citep{foreman13} to determine our best fit model parameters and their
posterior probabilities.  This code implements the affine-invariant
ensemble sampler of \citet{goodman10}.  It generates several MCMC
chains (``walkers'') that evolve through the model space
simultaneously, with different model parameters (represented by
$\overrightarrow\theta$ below) each time step.  We typically use 86
walkers here.  We do not use the first $\approx 2000$ states for each
walker (slightly different for each model) to avoid the ``burn-in''
period.

The MCMC algorithm uses a likelihood function, $\mathcal{L} \propto
\exp(-\chi^2)$.  For most of our models, we include fits to the
$\g$-ray opacity ($\tau$), luminosity density ($j$), stellar
mass density ($\rho$), EBL intensity ($I_{\rm EBL}$), and FUV dust
extinction ($A_{\rm FUV}$) data described above.  Thus,
\begin{flalign}
\chi^2 & = \sum_i \frac{ [\tau_{\rm obs}(E_i,z_i) - \tau_{\rm
      model}(\overrightarrow\theta| E_i,z_i)]^2}{\sigma_{\tau,i}^2} 
 + \sum_i \frac{ [j_{\rm obs}(\lambda_i,z_i) - j_{\rm
      model}(\overrightarrow\theta| \lambda_i,z_i)]^2}{\sigma_{j,i}^2} 
\nonumber \\ &
 + \sum_i \frac{ [\rho_{\rm obs}(z_i) - \rho_{\rm
      model}(\overrightarrow\theta| z_i)]^2}{\sigma_{\rho,i}^2} 
 + \sum_i \frac{ [I_{\rm EBL,obs}(\lambda_i,z_i) - I_{\rm EBL,
      model}(\overrightarrow\theta| \lambda_i,z_i)]^2}{\sigma_{I_{\rm EBL},i}^2} 
\nonumber \\ &
 + \sum_i \frac{ [A_{\rm FUV,obs}(z_i) - A_{\rm FUV,
      model}(\overrightarrow\theta| z_i)]^2}{\sigma_{A_{\rm FUV},i}^2} 
\ .
\end{flalign}
In some cases the upper and lower errors differ. In these cases, if
$\tau_{\rm obs}(E_i,z_i)>\tau_{\rm model}(\overrightarrow\theta|
E_i,z_i)$ we use the lower error for $\sigma_{\tau,i}$; otherwise we
use the upper error.  We similarly discriminate between upper and
lower errors for the other data points.

We use flat priors for all our parameters with appropriately large
limits.  Our results are generally independent of the limits on the
priors except for the fit to the $\tau_{\g\g}$ only (Model B) as
described by \citet{abdollahi18}.  

\section{Results}
\label{results}

The median and 68\% confidence intervals of the posterior
probabilities of the model parameters resulting from all of our MCMC
fits are reported in Table \ref{modeltable}.  The table also
summarizes the differences in the models, including the IMF and SFR
parameterization used, the data sets used, and the parameters that
were fixed or free during the model fits.

\begin{deluxetable*}{lccccccc}
\tablecaption{Model parameters.  }
\tablewidth{0pt}
\tablehead{
\colhead{} &
\colhead{Model A} & 
\colhead{Model A.c} & 
\colhead{Model B} & 
\colhead{Model C} & 
\colhead{Model D} & 
\colhead{Model D.c} & 
\colhead{Model E}}  
\startdata
IMF\tablenotemark{$\dagger$}       & BG03  & BG03 & BG03         & BG03     & Sal  & Sal   & BG03 \\
SFR\tablenotemark{$\dagger\dagger$}       & MD14  & MD04 & MD14         & MD14     & MD14  & MD14  & piece \\
dust      & free  & free & fixed        & free     & free  & free  & free \\
cosmology & fixed & free & fixed        & fixed    & fixed & free & fixed \\
\hline
data\tablenotemark{$\ddagger$}      & all  & all        & $\tau_{\g\g}$ & $j+\rho$ & all   & all   & all \\
\hline
$a_s\ [10^{-3}]$ & $9.2^{+0.5}_{-0.4}$ & $9.1^{+0.5}_{-0.6}$   & $9.6^{+7.3}_{-5.2}$  & $9.4\pm0.4$        & $13.3\pm0.6$        & $14.5^{+1.2}_{-0.8}\pm0.7$ & $-2.04\pm0.02$ \\
$b_s$         & $2.79^{+0.10}_{-0.09}$ & $2.63\pm0.14$ & $3.1^{1.4}_{-1.0}$  & $2.73^{+0.11}_{-0.10}$ & $2.61^{+0.10}_{-0.09}$ & $2.71\pm0.12$ & $ 2.81\pm0.11$\\ 
$c_s$         & $3.10^{+0.10}_{-0.09}$ & $3.22^{+0.10}_{-0.12}$ & $3.0^{+0.7}_{-0.5}$     & $3.11^{+0.11}_{-0.09}$ & $2.91\pm0.08$ & $2.95^{+0.11}_{-0.09}$ & $1.25\pm0.25$ \\ 
$d_s$         & $6.97^{+0.16}_{-0.15}$ & $6.89^{+0.22}_{-0.16}$  & $8.0^{+1.4}_{-1.5}$ & $6.91\pm0.18$  & $6.03\pm0.09$ & $6.26^{+0.19}_{-0.22}$ & $-1.25^{+0.60}_{-0.57}$ \\
$e_s$         & N/A                & N/A                & N/A & N/A & N/A & N/A & $-1.84^{+0.74}_{-0.70}$ \\
$f_s$         & N/A                & N/A                & N/A & N/A & N/A & N/A & $-4.40^{+0.19}_{-0.23}$  \\
$f_{esc,1}$    & $1.88^{+0.24}_{-0.44}$ & $1.83^{+0.42}_{-0.53}$ & $0.257$\tablenotemark{*}  & $1.96^{+0.23}_{-0.39}$ & $1.94^{+0.25}_{-0.44}$ & $0.68^{+0.52}_{-0.37}$ & $1.93^{+0.22}_{-0.42}$ \\
$f_{esc,2}$    & $2.18^{+0.29}_{-0.51}$ & $2.08^{+0.54}_{-0.57}$ & $0.287$\tablenotemark{*}  & $2.58^{+0.36}_{-0.51}$ & $2.19^{0.30}_{-0.49}$ & $0.77^{+0.64}_{-0.42}$ & $2.26^{+0.26}_{-0.49}$ \\
$f_{esc,3}$    & $2.93^{+0.38}_{-0.70}$ & $2.73^{+0.51}_{-0.82}$ & $0.271$\tablenotemark{*}  & $2.85^{+0.36}_{-0.58}$ & $2.99^{+0.40}_{-0.70}$ & $1.01^{+0.79}_{-0.52}$ & $3.00^{+0.37}_{-0.65}$ \\
$f_{esc,4}$    & $3.93^{+0.49}_{-0.93}$ & $3.68^{+0.77}_{-1.12}$ & $0.628$\tablenotemark{*}  & $4.00^{+0.42}_{-0.82}$ & $3.62^{+0.47}_{-0.81}$ & $1.20^{+0.87}_{-0.61}$ & $4.01^{+0.43}_{-0.87}$ \\
$f_{esc,5}$    & $8.57^{+1.07}_{-2.00}$ & $8.16^{+1.51}_{-2.68}$ & $0.959$\tablenotemark{*}  & $8.73^{+0.92}_{-1.76}$ & $8.50^{+1.07}_{-1.93}$ & $2.91^{+2.21}_{-1.55}$ & $8.80^{+0.88}_{-1.89}$ \\
$m_d$         & $1.52^{+0.04}_{-0.04}$ & $1.43^{+0.04}_{-0.03}$ & $1.49$\tablenotemark{*}  & $1.59\pm0.05$       & $1.51\pm0.04$ & $1.47^{+0.05}_{-0.04}$ & $1.53\pm0.04$ \\
$n_d$         & $0.35^{+0.04}_{-0.04}$  & $0.41^{+0.05}_{-0.04}$ & $0.644$\tablenotemark{*} & $0.35\pm0.04$       & $0.17\pm0.03$ & $0.22\pm0.05$ & $0.37^{+0.06}_{-0.05}$ \\
$p_d$         & $4.12^{+0.13}_{-0.13}$ & $4.14^{+0.15}_{-0.16}$ & $3.4$\tablenotemark{*}   & $4.15^{+0.12}_{-0.13}$ & $3.86^{+0.13}_{-0.12}$ & $3.91^{+0.13}_{-0.18}$ & $4.71^{+0.32}_{-0.39}$ \\
$q_d$         & $5.89^{+0.55}_{-0.48}$ & $5.12^{+0.36}_{-0.57}$ & $3.54$\tablenotemark{*}  & $6.11^{+0.54}_{-0.48}$ & $6.95^{+0.75}_{-0.66}$ & $5.97^{+0.66}_{-0.57}$ & $3.67^{+0.70}_{-0.57}$ \\
$f_1$         & $0.56^{+0.17}_{-0.18}$ & $0.13^{+0.20}_{-0.06}$  & $0.48^{+0.24}_{-0.30}$ & $0.60^{+0.14}_{-0.18}$ & $0.55^{+0.16}_{-0.17}$ & $0.19\pm0.11$ & $0.59^{+0.15}_{-0.19}$ \\
$f_2$         & $0.26^{+0.18}_{-0.17}$ & $0.68^{+0.06}_{-0.21}$  & $0.47^{+0.32}_{-0.26}$ & $0.22^{+0.18}_{-0.14}$ & $0.25^{+0.17}_{-0.16}$ & $0.63^{+0.10}_{-0.12}$ & $0.23^{+0.20}_{-0.15}$ \\
$T_1$ [K]     & $60.5^{+2.3}_{-3.5}$ & $40.5^{+11.7}_{-4.2}$ & $40$\tablenotemark{*} & $61^{+2}_{-3}$ & $60.8^{+2.1}_{-3.2}$ & $62.8^{+2.1}_{-3.1}$ & $51.9^{+6.7}_{-10.6}$ \\
$H_0\ [\km\ \s^{-1}\ \Mpc^{-1}]$ & $70$\tablenotemark{*} & $69.8^{+3.6}_{-3.2}$ & $70$\tablenotemark{*} & $70$\tablenotemark{*} & $70$\tablenotemark{*}  & $79.4^{+8.1}_{-4.3}$ & $70$\tablenotemark{*} \\ 
$\Omega_m$    & $0.3$\tablenotemark{*} & $0.23^{+0.3}_{-0.4}$ & $0.3$\tablenotemark{*} & $0.3$\tablenotemark{*} & $0.3$\tablenotemark{*} & $0.29\pm0.04$ & $0.3$\tablenotemark{*}  
\enddata
\tablenotetext{*}{Parameters fixed in the fit.}
\tablenotetext{$\dagger$}{Initial Mass function.  
Sal = \citet{salpeter55}, Equation (\ref{SalpeterIMFeqn}).  
BG = \citet{baldry03}, Equation (\ref{IMFeqn}).  }
\tablenotetext{$\dagger\dagger$}{Star Formation Rate parameterization.  
MD14 = \citet{madau14}, Equation (\ref{SFRmadau}).  
piece = piecewise parameterization, Equation (\ref{piecewiseSFReqn}).  }
\tablenotetext{$\ddagger$}{The data fit with the model.}
\label{modeltable}
\vspace{2mm}
\end{deluxetable*}

\subsection{Fiducial Model Comparison with Data}
\label{fiducialcompare}

We consider ``Model A'' to be our fiducial model.  This model uses the
\citet{madau14} SFRD parameterization, Equation (\ref{SFRmadau}), and
the \citet{baldry03} IMF, Equation (\ref{IMFeqn}), and has all of the
SFRD and dust extinction and emission parameters free in the fit, but
keeps the cosmological parameters fixed.  It fits all of the data
described in Section \ref{data}: the $\g$-ray opacity, luminosity
density, stellar mass density, and FUV dust extinction data, as
well as the EBL intensity lower limits.

\begin{figure}
\vspace{2.2mm} 
\epsscale{1.1} 
\plotone{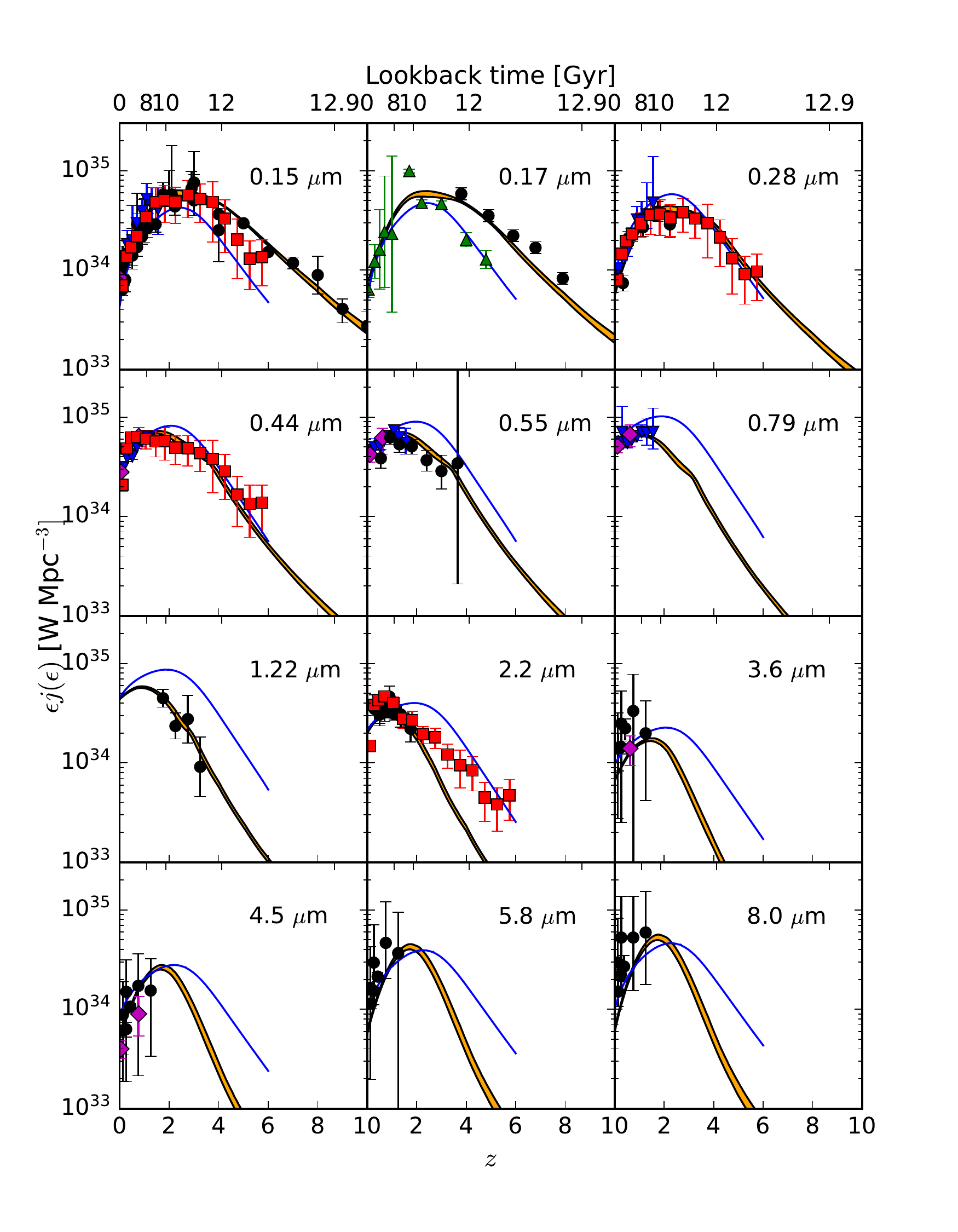}
\caption{The 68\% contour luminosity density result for Model A as a
  function of $z$ at twelve wavelengths (shaded orange region).  The red
  squares are the data from \citet{saldana21}, the violet dimonds
  represent the data from \citet{andrews17}, and the blue
  downward-pointing triangles represent the data from
  \citet{tresse07}.  The rest of the data from galaxy surveys are
  plotted as the black circles.  All of these data were used in the
  model fit, while the green upward-pointing triangles represent
  older, superseded data from \citet{sawicki06} that were not used in
  the fit.  The blue curves are the luminosity densities from the
  older Model C from \citet{finke10_EBL}.  The wavelengths of the data
  and models are labeled on the plots.  }
\label{LD_modelA}
\vspace{2.2mm}
\end{figure}

The luminosity densities as a function of $z$ from our model are
plotted in Figure \ref{LD_modelA}, and compared with data.  The model
is clearly a good fit to the most recent data at these wavelengths.
It is interesting to note the divergence at high $z$ with 
  the older data and model.  In the FUV band, the $0.17\ \mu\m$
older data from \citet{sawicki06} used in the previous model by
\citet{finke10_EBL} is plotted.  \citet{sawicki06} used deep imaging
with the Keck telescope.  The updated deep {\em HST} observations
\citep{bouwens16} have found much more light at these wavelengths.
Since the model from \citet{finke10_EBL} was designed to reproduce
these observations (among others), it is not surprising that it is
lower than Model A of this work.  One should note that the models from
\citet{finke10_EBL} did not extend beyond $z=6$.  At $0.28\ \mu\m$ and
$0.44\ \mu\m$, the older model, current model, and all data are in
reasonably good agreement.

As one moves to longer wavelengths, our Model A becomes lower than the
older model from \citet{finke10_EBL}, particularly at $z>2$.  The
model reproduces most of the data at most wavelengths shown here.  For
instance, the current Model A follows the data from \citet{stefanon13}
at $1.22\ \mu\m$, which was obviously not available to
\citet{finke10_EBL}.  However, our Model A is considerably below the
luminosity density measurements from \citet{saldana21} at
$2.2\ \mu\m$, particularly at $z\ga2$, despite the fact that these
data were included in the fit.  We think this is because of our poor
modeling of the PAH component (see below). The model is generally
consistent with the \citet{saldana21} data at $z\la2$ where the
emission is dominated by stellar emission, rather than PAH
emission. The model also closely follows the lower luminosity
densities of \citet{stefanon13} at $1.2\ \mu\m$ and $1.6\ \mu\m$ at
$z\la2$.  The stellar models seem to be highly constraining in this
wavelength range.  However, this is the wavelength range identified by
\citet{conroy09} where assumptions about all stars having the same
metallicity, rather than a distribution of metallicities, can create
uncertainty in emission.  This is illustrated in Figure
\ref{ssp_compare} where different metallicities can create significant
differences at the redder wavelengths.  At the longer IR wavelengths
($3.6\ \mu\m$ to $8.0\ \mu\m$), most of the data are from the work of
\citet{babbedge06} and \citet{dai09}, and our model fits them well.
However, at these wavelengths, the data do not extend to very high
redshift, and the data that do exist have large uncertainties; thus
this wavelength range is not strongly constrained.  This paucity of
data should be rectified in the near future with the {\em James Webb
  Space Telescope} Mid-Infrared Instrument.  At $3.6$ -- $8.0\ \mu\m$
and $z\ga2$ our model is considerably lower than the older model from
\citet{finke10_EBL}.

\begin{figure}
\vspace{2.2mm} 
\epsscale{1.25} 
\plotone{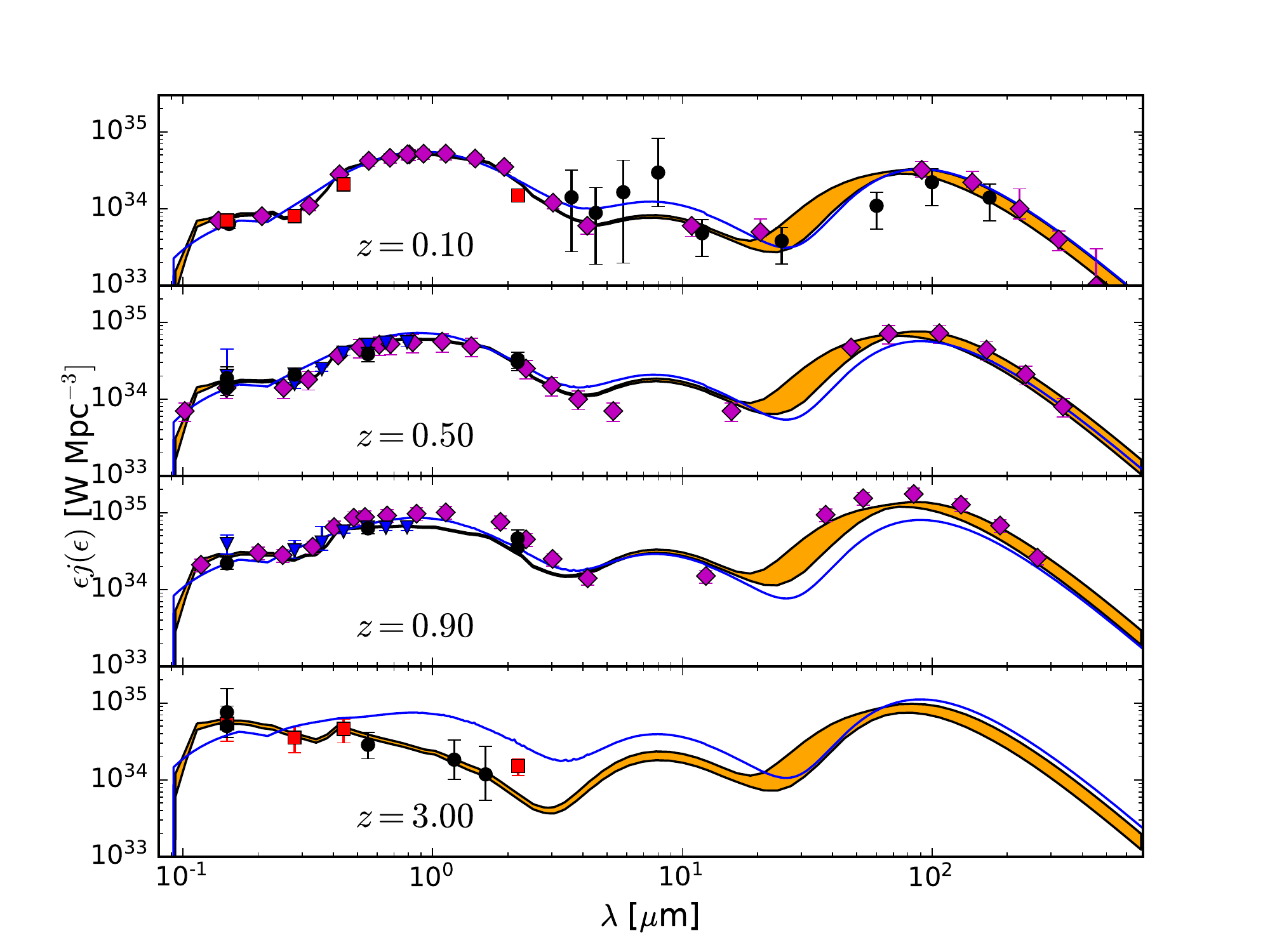}
\caption{The 68\% contour luminosity density result for Model A as a
  function of $\lambda$ at four redshifts (shaded orange region).  The
  red squares are the data from \citet{saldana21}, the violet dimonds
  represent the data from \citet{andrews17}, and the blue
  downward-pointing triangles represent the data from
  \citet{tresse07}.  The rest of the data from galaxy surveys used in
  the fit are plotted as the black circles.  The blue curves are the
  luminosity densities from the older Model C from
  \citet{finke10_EBL}.  The redshifts of the data and models are
  labeled on the plots.}
\label{LD_modelA_staticz}
\vspace{2.2mm}
\end{figure}

The luminosity density as a function of wavelength at four redshifts
is shown in Figure \ref{LD_modelA_staticz}.  At low redshifts ($z\le
0.5$) the model follows the data from \citet{andrews17} closely.  At
$z=0.10$, older measurements at 3.6 -- 8 $\mu\m$ by \citet{babbedge06}
are generally higher but also have greater uncertaintly compared with
more recent measurements by \citet[][see also large uncertainties in
  these data in Figure \ref{LD_modelA}]{andrews17}.  At this
redshift, our updated Model A here and the model from
\citet{finke10_EBL} are in agreement at all wavelengths except the
$\lambda\approx 3.0$ -- $20\ \mu\m$ range, mainly due to the influence
of the \citet{babbedge06} data on the older model.  For the $\ge
12\ \mu\m$ measurements by \citet{takeuchi06}, the measurements are a
bit lower than the model and other measurements, but still mostly
consistent with them, considering the larger uncertainties.  At
$z=0.5$, essentially all the data and models are in agreement.

At $z=0.9$, some separation between models and data can be seen.  At
this redshift, at $\lambda < 1.2\ \mu\m$, the measurements from
\citet{tresse07} are below the ones from \citet{andrews17}.  Our
Model A follows the \citet{tresse07} measurements here more closely,
which is quite a bit lower than Model C from \citet{finke10_EBL} in
the $\lambda = 0.8$ -- $3.0\ \mu\m$ range.  At $\lambda>30\ \mu\m$, the
updated model is higher than the older model from \citet{finke10_EBL}.
This is due to the influence of the \citet{andrews17} data at these
wavelengths.  This redshift ($z=0.9$) is the highest redshift 
for which \citet{andrews17} have measurements.

The luminosity density at $z=3$ is a particularly interesting case.
The older model from \citet{finke10_EBL} is quite divergent from our
updated model here.  In the model itself, this seems to be related to
the updated stellar spectra. The data we fit is also influencial, and
the updated model reproduces the data at $\lambda<3\ \mu\m$ well.
Most of the optical-near IR measurements used at this redshift
\citep{marchesini12,stefanon13,saldana21} did not exist when
\cite{finke10_EBL} created their model.  The only data point at $z=3$
that is not reproduced by the model is the $\lambda=2.2\ \mu\m$
measurement by \citet{saldana21}.  This could be due to the
approximation of the PAH component by a blackbody.  A more realistic
treatment of this component might reproduce the high redshift
$2.2\ \mu\m$ data better, where their is considerable disagreement
between data and model (see Figure \ref{LD_modelA}).  Another
possibility is that out model is underestimating the 
metallicity (see Figure \ref{massmetallicity_modelA}), and that
a higher metallicity would lead to more intense luminosity density
at this wavelength (Figure \ref{ssp_compare}).  Finally, it could be
that AGN, neglected
here, could contribute to the $2.2\ \mu\m$ luminosity density
observed by \citet{saldana21}.  The AGN would be required to make up
70\% of the $2.2\ \mu\m$ luminosity density at $z=3.75$, and 90\% at
$z=5.75$.  Estimates of the AGN contribution to the EBL indicate that
it likely makes no more than 10\% at any relevant wavelength
\citep{dominguez11,andrews18,abdollahi18,khaire19}.  We therefore
believe it unlikely that AGN could be the reason for the discrepancy
here. 

\begin{figure}
\vspace{2.2mm} 
\epsscale{1.2} 
\plotone{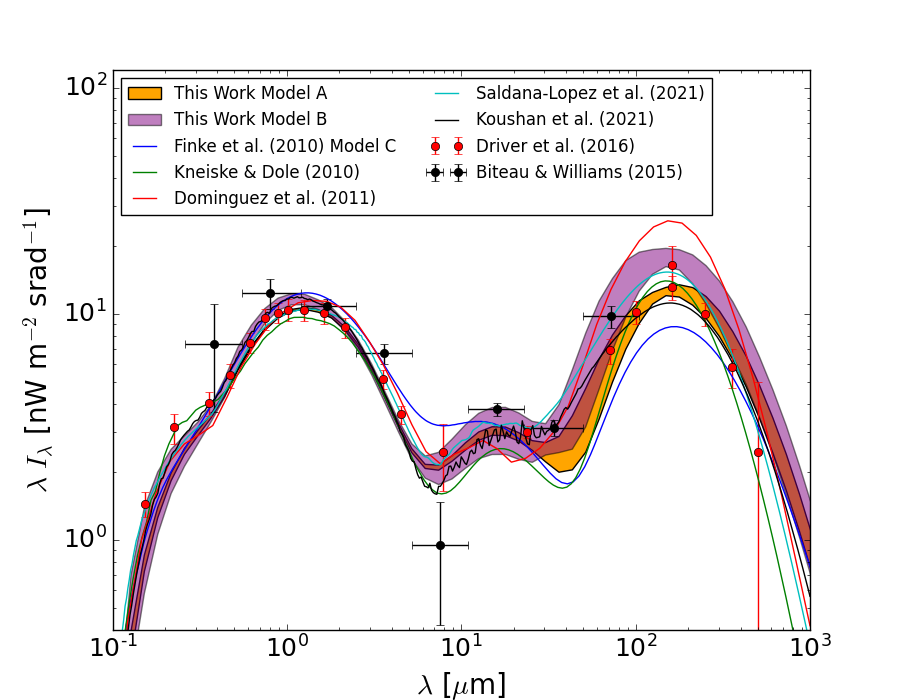}
\caption{The 68\% contour EBL intensity results as a function of
  $\lambda$ at $z=0$ for Models A and B.  A variety of lower limit
  data and models curves are shown as well, as seen in the legend.}
\label{EBLintensity}
\vspace{2.2mm}
\end{figure}

The 68\% confidence interval for the EBL intensity at $z=0$ for our
Model A is shown in Figure \ref{EBLintensity}, along with a variety of
lower limits and other models from the literature.  Our model result
has little uncertainty at $\lambda \la 10\ \mu\m$, while the
uncertainty grows at longer wavelengths.  Our updated model's EBL
intensity is lower than most other models at $\lambda \la 10\ \mu\m$,
and quite close to being in conflict with some of the lower limits
from \citet{driver16}, and below the ones from \citet{biteau15}.  The
only lower EBL model in this wavelength range is the one from
\citet{kneiske10}, which was designed to be the lowest possible model
consistent with lower limits.  At longer wavelengths, uncertainty is
greater.  The far-IR peak at $\approx 200\ \mu\m$ is greater than the
older model from \citet{finke10_EBL}, considerably lower than the one
from \citet{dominguez11}, and mostly consistent with the others.

\begin{figure}
\vspace{2.2mm} 
\epsscale{1.25} 
\plotone{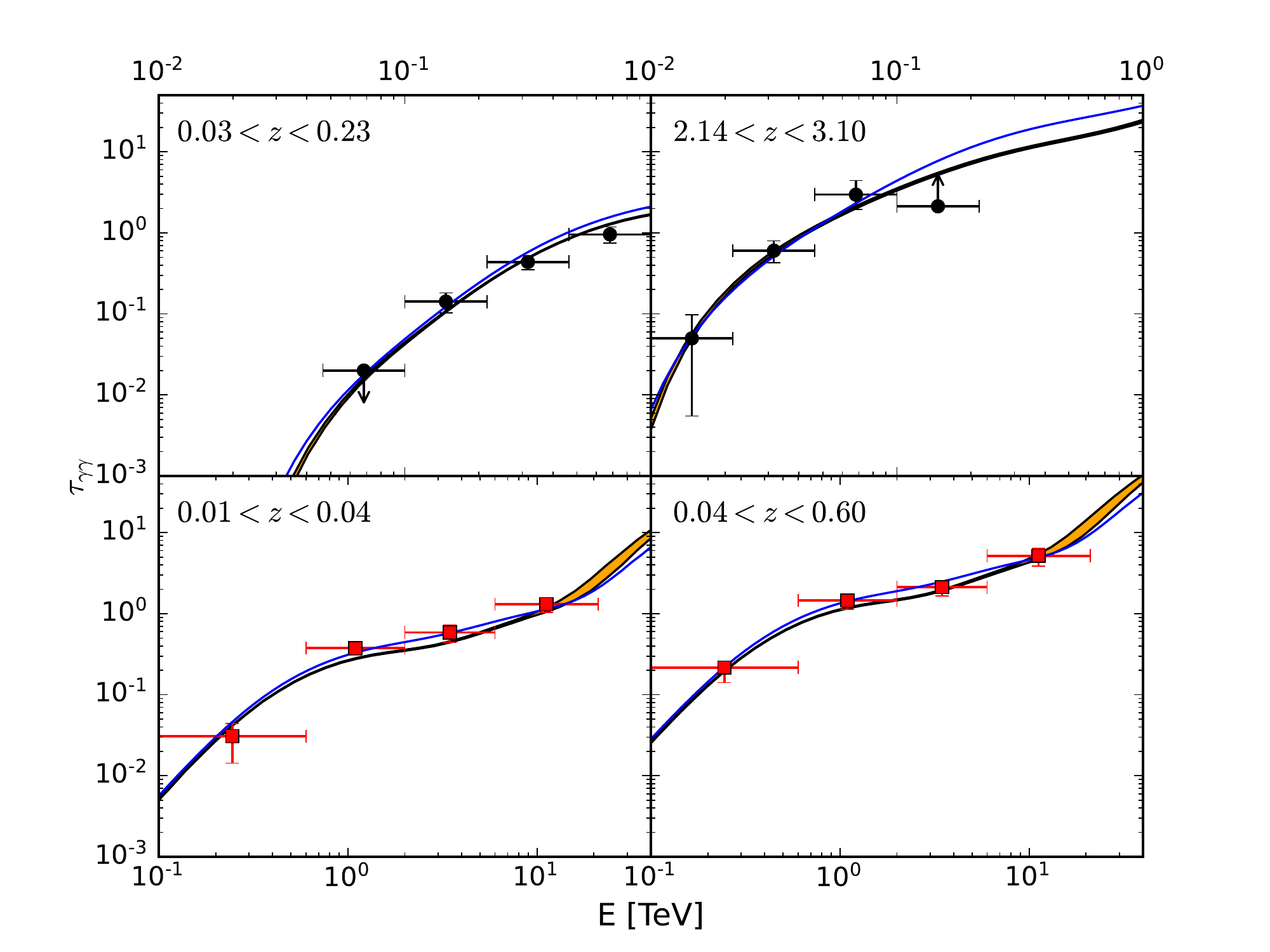}
\caption{The 68\% contour EBL absorption optical depth as a function
  of observed energy for different redshift ranges for our Model A.
  The LAT absorption measurements from \citet{abdollahi18} and the
  IACT measurement from \citet{desai19} are shown as the black circles
  and red squares, respectively.  The blue curve is the older Model C
  from \citet{finke10_EBL}.  Note the different energy scales on the 
  different plots.}
\label{tauplot}
\vspace{2.2mm}
\end{figure}

The EBL absorption optical depth from our Model A for four redshift
ranges is shown in Figure \ref{tauplot}.  These are compared with the
EBL absorption measurements with the {\em Fermi}-LAT
\citep{abdollahi18} and IACTs \citep{desai19}, and the previous best
fit model from \citet{finke10_EBL}.  In all cases, the 68\% model
errors on the absorption optical depth are quite small, typically $\la
10\%$. The updated model is in quite good agreement with the data and
with the previous 2010 model in almost all cases.  In the lowest LAT
redshift bin, $0.03<z<0.23$, the new model is $\approx 20\%$ lower
than the old one at most energies.  At the highest LAT redshift bin,
$2.14<z<3.10$, agreement between the old and new model is good below
$0.1\ \TeV$, although above this energy the old model predicts
$\approx 50\%$ higher $\tau_{\g\g}$.  Measurements of $\tau_{\g\g}$
are not found in this redshift range for energies $>0.1\ \TeV$, since
the values of $\tau_{\g\g}$ are likely too high here to be measured 
(see Section \ref{graydatasection}).

IACTs are sensitive to higher energy $\g$-rays than the LAT, which
makes them sensitive to $\g$-ray absorption at lower redshifts and
higher energies.  The absorption measurements with IACTs are shown in
the bottom two panels of Figure \ref{tauplot}, along with the
comparison to our updated model and the \citet{finke10_EBL}
model. Agreement between data and both the models is very good in most
places.  The data and the older model are $\approx 30\%$ higher in the
1 -- 6 TeV range than the updated model for both redshift bins.  In this
energy range, the $\g$-rays interact with EBL photons in the range
$\approx 4-24\ \mu\m$; see Equation (\ref{energyeqn}).  In this range,
the EBL photons are due to the red
end of the stellar photons, and the PAH component, where the old
model from \citet{finke10_EBL} is clearly below the updated model
(Figure \ref{EBLintensity}).  

The Model A fit is mostly constrained by the luminosity density
measurements, since these are more plentiful and have smaller errors.
So the discrepancy in the model and observed $\g$-ray absorption here
could be interpreted as a discrepancy between the luminosity density
and $\g$-ray absorption measurements.

\subsection{Physical Implications}
\label{physimplication}

\begin{figure}
\vspace{2.2mm} 
\epsscale{1.17} 
\plottwo{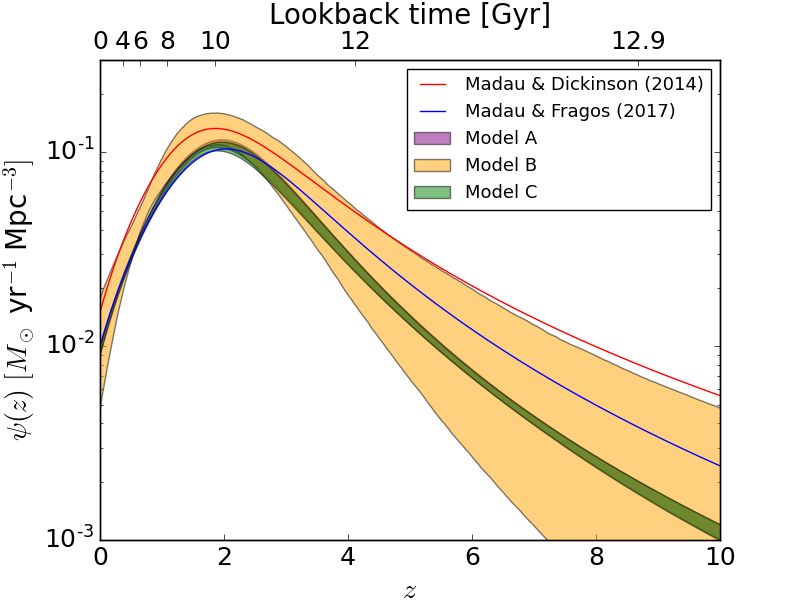}{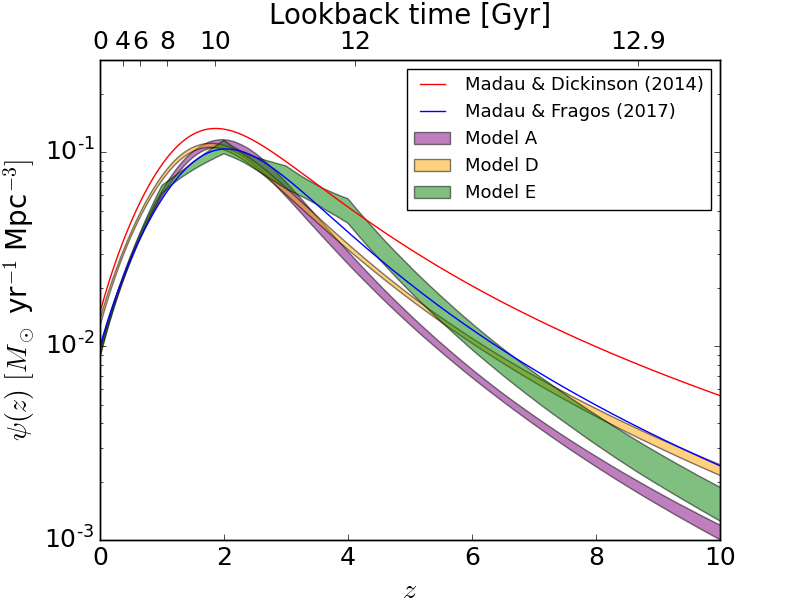}
\caption{The 68\% contour of the SFRD, $\psi(z)$, as a function of
  redshift $z$ for our models, as described in the legend.  Also shown
  are the SFRD from \citet{madau14} and \citet{madau17}.
    Models A and C are nearly indistinguishable on the left plot.}
\label{SFR_modelABC}
\vspace{2.2mm}
\end{figure}

The SFRD for the result of Model A is shown in Figure
\ref{SFR_modelABC}.  The 68\% SFR shows little uncertainty, rarely
more than 10\%.  The curve from \citet{madau17} is within our model
uncertainty at $z\la3$. But compared with \citet{madau14}, our model
predicts a lower SFRD by a factor $\approx 2$ at $z\la3$, and by a
factor $\approx 5$ at $z\ga3$.  The likely discrepancy between our
model and \citet{madau14} is that the latter used a Salpeter IMF,
while we used the Baldry-Glazebrook IMF.  \citet{madau17} used the
Kroupa IMF \citep{kroupa01} which is more similar to the IMF we used,
in that it deviates from a power-law at low masses, with fewer stars.
The discrepancy at $z\la3$ is likely due to different assumptions
about the evolution of metallicity and dust extinction with redshift.

\begin{figure}
\vspace{2.2mm} 
\epsscale{1.17} 
\plottwo{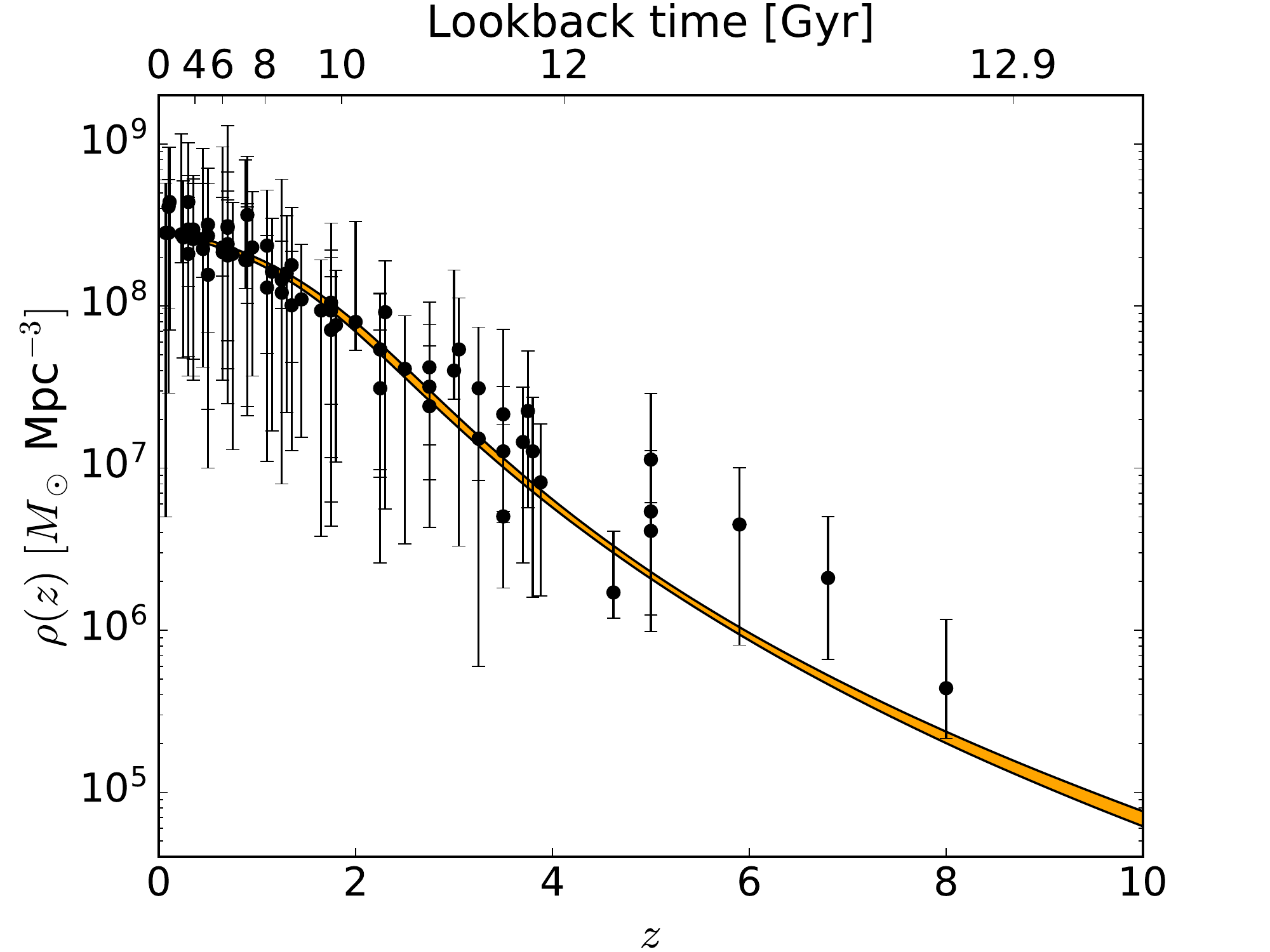}{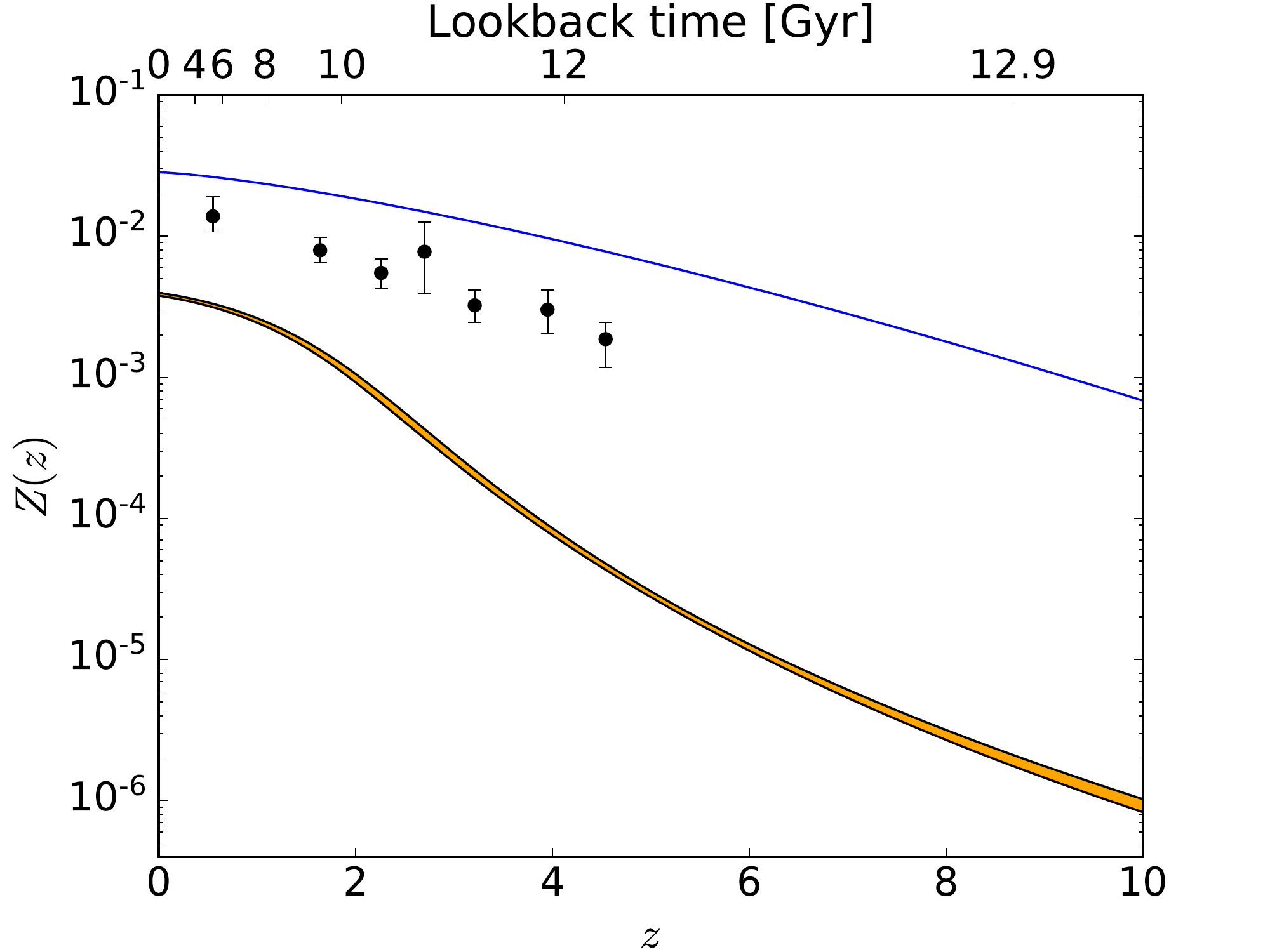}
\caption{{\em left}: The 68\% confidence interval (orange countour) on
  the stellar mass density as a function of $z$ from Model A,
  along with data (black circles) from \citet{madau14}.  {\em right}:
  68\% confidence interval (orange countour) on the gas-phase
  metallicity as a function of $z$ from Model A, along with the fit
  from \citet[][blue curve]{madau17}.  Mean metallicity
    measurements from damped Ly$\alpha$ systems corrected for dust
    depletion \citep{peroux20} are shown as black circles.}
\label{massmetallicity_modelA}
\vspace{2.2mm}
\end{figure}

The resulting stellar mass density and gas-phase
metallicity as a function of $z$ are shown in Figure
\ref{massmetallicity_modelA}.  The model mass density reproduces the
data quite well.  Metallicity measurements from damped
  Ly$\alpha$ systems \citep[][and references therein]{peroux20}
  corrected for dust depletion following \citet{decia18} are also
  shown on the figure.  Our model is clearly below these data points.
  We have also performed a model fit identical to Model A, but also
  including a fit to these points. The result is nearly identical; the
  metallicity data do not have enough constraining power to impact the
  model fit.  The metallicities from Ly$\alpha$ systems compiled by
  \citet{peroux20} \citep[and also][]{biffi13} have large dispersions
  around the mean.  This can also be seen in hydrodynamic chemical
  evolution simulations of early galaxy star formation; these
  simulations indicate galaxies at $z\approx 9$ have $10^{-9} < Z <
  10^{-4}$ \citep{biffi13}.  The model mean metallicity is compared
  with the mean metallicity curve from \citet[][]{madau17}.  For all
  metallicity conversions in this figure we assumed
  $Z_\odot=0.02$. The \citet{madau17} curve was found by
convolving the galaxy mass-metallicity relation for star-forming
galaxies with various galaxy mass functions at different redshifts.
Their results are clearly discrepant with our model, where they get
much higher metallicities by $\approx 1.5-2.5$ orders of magnitude.
It is not clear if these two methods are really measuring the same
thing. The \citet{madau17} curve is measuring the
  galaxy mass-weighted mean gas phase metallicity, while the
  interpretation of our mean metallicity from a one-zone closed box
  model is not clear.  The mass-metallicity relation used by
  \citet{madau17} has been derived from only star-forming galaxies at
  $z\la1.6$, so applying it to all galaxies may not be appropriate,
  nor using it at $z\ga1.6$.  The \citet{madau17} curve is also above
  almost all of the damped Ly$\alpha$ data points, even when corrected
  for dust depletion.  \citet{madau17} also note the metallicity
  normalization of the relation is quite uncertain.  It may also be
  that the closed box model we use is not a good approximation for the
  metallicity evolution of the univese; indeed it is almost certainly
  an over-simplification.  

\begin{figure}
\vspace{2.2mm} 
\epsscale{1.0} 
\plotone{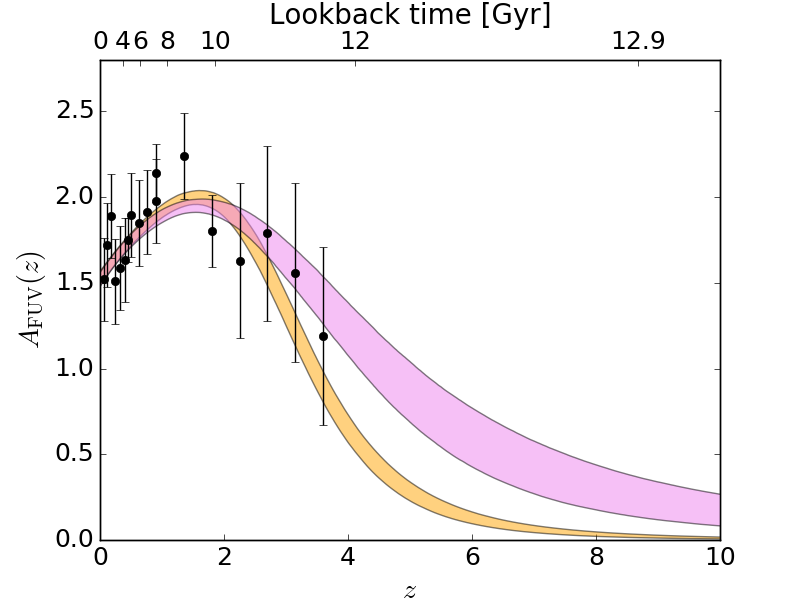}
\caption{The 68\% contour FUV dust extinction, $A(z)$, as a function
  of redshift for Model A (orange shaded region) and Model E (violet
shaded region).  Observational data are shown as black
  circles.}
\label{dustext_fig}
\vspace{2.2mm}
\end{figure}

The 68\% contour of our Model A FUV (1500 \AA) dust extinction as a
function of redshift (Equation [\ref{UVabscoeff}]) is shown in Figure
\ref{dustext_fig}.  It clearly agrees
with the data.  The extinction increases with increasing redshift
until about $z=2$, after which it decreases.  As $z\rightarrow10$, 
$A_{\rm FUV}(z)\rightarrow0$, as expected; since dust is made of metals,
and at high redshift the metallicity $Z\rightarrow0$.  

\begin{figure}
\vspace{2.2mm} 
\plotone{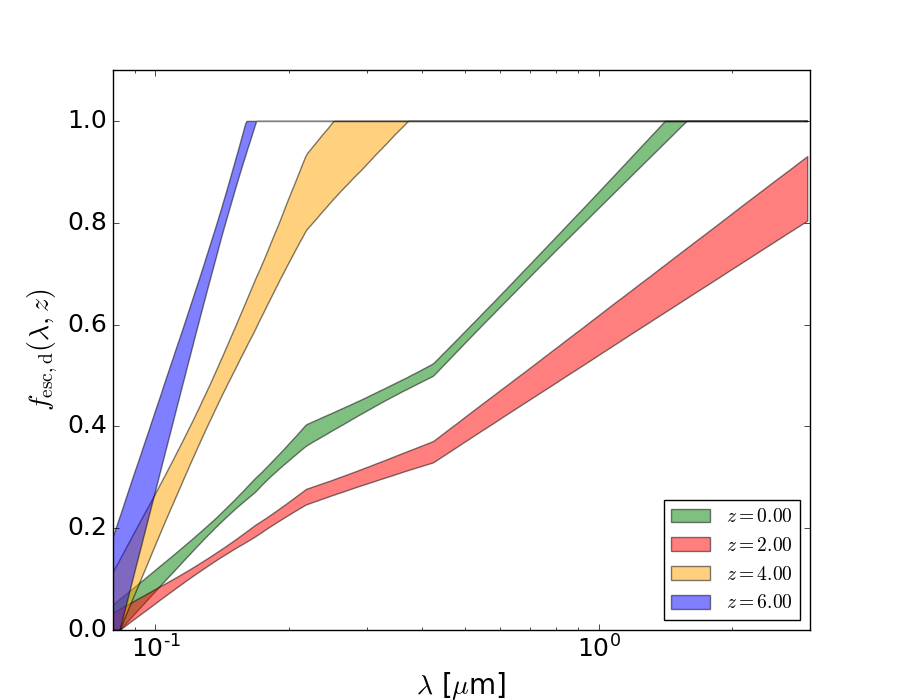}
\caption{The 68\% contour of the dust escape fraction, $f_{d,esc}(\lambda,z)$
for Model A.}
\label{fescd_fig}
\vspace{2.2mm}
\end{figure}

Not only does $A(z)$, i.e. the normalization of the dust extinction
curve, evolve, the shape of the curve evolves as well.  The dust
escape fraction, $f_{\rm esc,d}(\lambda,z)$ is shown in Figure
\ref{fescd_fig} as a function of $\lambda$ for four different
redshifts.  The general pattern is that as the overall absorption decreases
(i.e., more photons escape), it decreases at longer
wavelengths much more than shorter wavelengths.  At the highest
redshift, $z=6$, above $\approx0.13\ \mu\m$, the escape fraction goes
to unity, indicating all the photons escape and none are absorbed.
One can also see in this figure that the escape fraction decreases
(absorption increases) until a peak at about $z=2$, and the escape
fraction increases (absorption decreases) at increasing redshift at
$z>2$.  This is consistent with what is seen in Figure
\ref{dustext_fig}.

\subsection{Contribution to local EBL}
\label{buildupsection}

\begin{figure}
\vspace{2.2mm}
\epsscale{1.17}
\plottwo{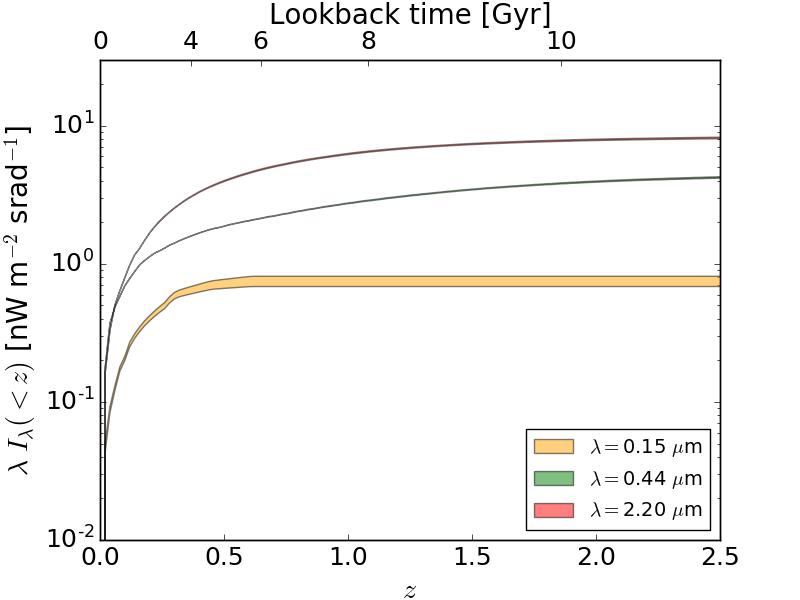}{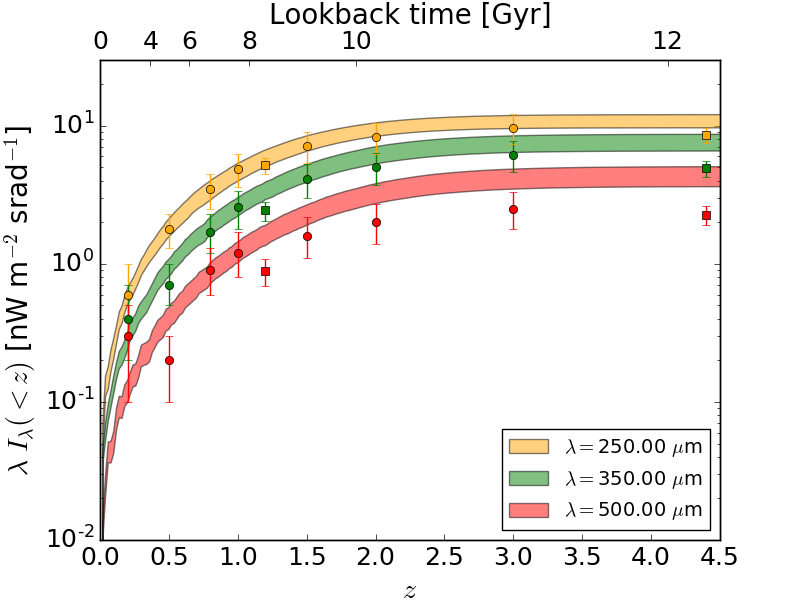}
\caption{The 68\% contours for the contribution to the local ($z=0$)
  EBL that originates at redshift $<z$ for various wavelengths from
  Model A, as shown in the legends.  The symbols represent the
  resolved contribution from galaxies, which are essentially {\em
    lower limits} on the contribution to the local EBL.  Square
  symbols are from BLAST \citep{marsden09} and circle symbols are from
  {\em Herschel}-SPIRE \citep{bethermin12}.}
\label{EBLbuildup}
\vspace{2.2mm}
\end{figure}

We can compute the contribution to the local ($z=0$) EBL intensity
from sources at redshifts $<z$ for our model as \citep{finke10_EBL}
\begin{flalign}
\e I(\e;<z) =  \frac{c}{4\pi} \int_0^z dz_1 \frac{ \ep j(\ep; z_1)}{1+z_1} 
\left|\frac{dt_*}{dz_1}\right|\ ,
\end{flalign}
where $\ep=\e(1+z_1)$.  The result of this calculation for Model A for
various wavelengths are shown Figure \ref{EBLbuildup}.  Also shown are
the resolved contributions to the build-up of the EBL, as measured by
BLAST \citep{marsden09} and {\em Herschel}-SPIRE \citep{bethermin12}.
These are lower limits on the contribution to the local EBL.  Based on
our modeling, most of the EBL has been resolved by {\em
  Herschel}-SPIRE at $250\ \mu\m$ and $350\ \mu\m$, while less has
been resolved at $500\ \mu\m$.  It is also apparent that at shorter
wavelengths for the stellar component of the EBL, the contribution to
the local EBL is much more ``local'' than for the sub-mm wavelengths.
Most of the local EBL comes from $z\la1$ for $\lambda<2.2\ \mu\m$, while
for the dust component in the sub-mm, one has to go out to
$z\approx2.5$ to have all of the local EBL.  This is consistent
with the result of \citet{saldana21}.

\subsection{Results with Different Data Sets}

Figure \ref{SFR_modelABC} also shows the resulting fit to the LAT
$\g$-ray absorption optical depth data only (Model B).  The addition
of luminosity and stellar mass density data significantly
increase the constraint on the SFRD.  Indeed, it is not clear if the
$\g$-ray absorption data is constraining the SFRD much more than the
luminosity and stellar mass density data.  To test this, we did
a run similar to Model A, fitting the luminosity, stellar mass
density, and dust extinction data, but without the $\g$-ray absorption
data.  This we called model C (Table \ref{modeltable}).  The results
are nearly identical, as seen in Figure \ref{SFR_modelABC}, with the
Model C result having slightly larger uncertainty ($\approx 5\%$) that
are nearly imperceptible in the figure.  It does not seem the $\g$-ray
absorption data is giving much more tightly constrained SFRD.

\begin{figure}
\vspace{2.2mm} 
\epsscale{1.1} 
\plotone{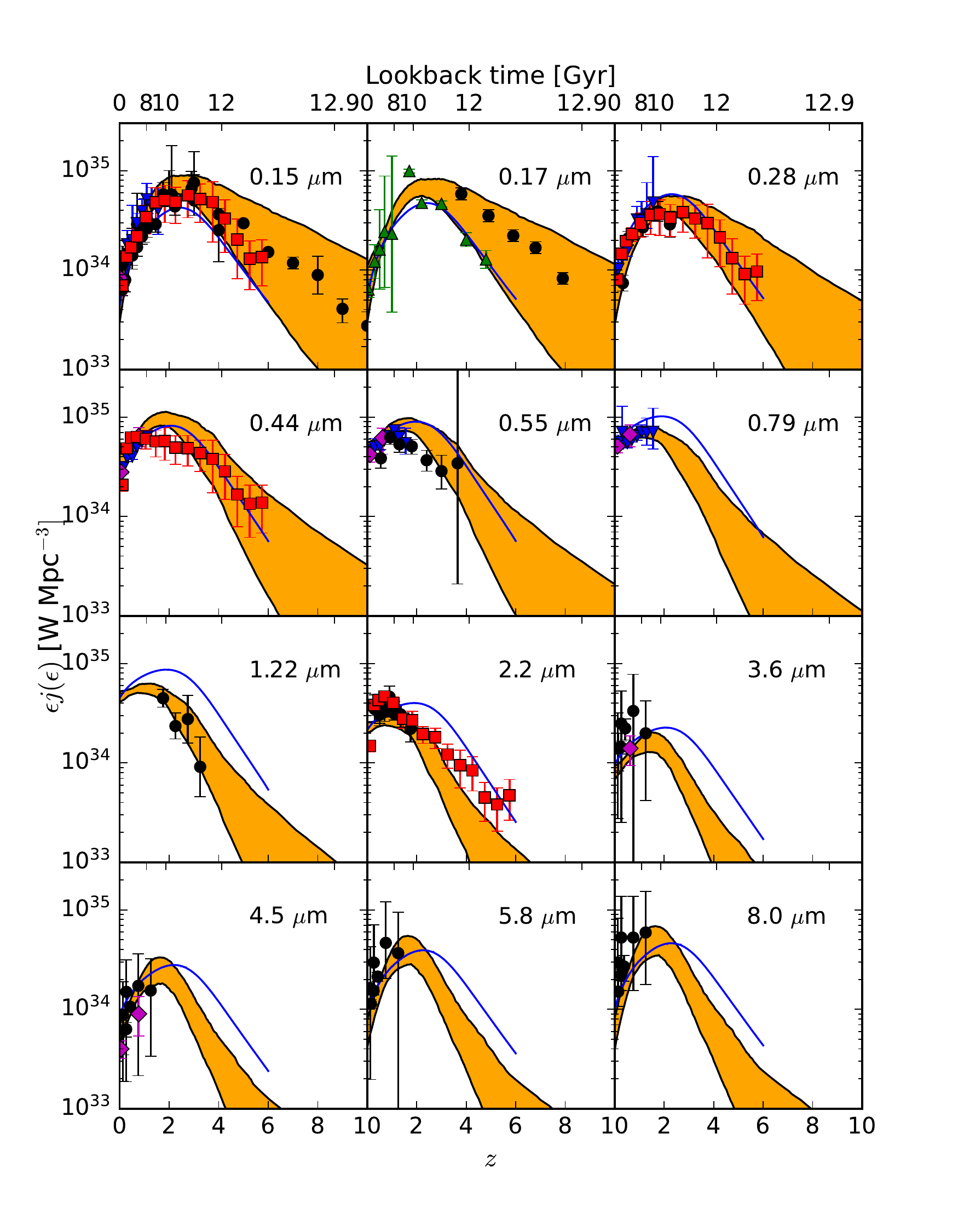}
\caption{The same as Figure \ref{LD_modelA} but the shaded regions are
  the 68\% confidence intervals are for model B, the fit to $\g$-ray
  data only.}
\label{lumdens_modelB}
\vspace{2.2mm}
\end{figure}

In Figure \ref{lumdens_modelB}, we plot the luminosity density model
results for the fit to the $\g$-ray absorption data only.  It
demonstrates that the $\g$-ray data alone can constrain the luminosity
density consistent with the observed luminosity density data.  Thus,
although the $\g$-ray absorption data do not contribute strongly to
the constraint due to their low $S/N$, they are valuable as an
alternative, independent measurement.

\subsection{Dust Emission}

The fractions of absorbed light being re-emitted by large warm grains
($f_1$), small hot grains ($f_2$), and polycyclic aromatic
hydrocarbons (PAHs) are constrained by our models that include
luminosity density (Table \ref{modeltable}), although uncertainties are
large.  We find results that are consistent with \citet{finke10_EBL}.
For the large warm grains, we find $f_1=0.56^{+0.17}_{-0.18}$,
compared with $f_1=0.6$ from the previous model; and for the small hot
grains we find $f_2=0.26^{+0.18}_{-0.17}$, a bit higher than the old
model ($f_2=0.05$), but within the uncertainty.

The results constrained by only the $\g$-rays (Model B) domonstrates
that the existing $\g$-ray data does not strongly constrain the dust
parameters, although the results of this model are completely
consistent with Model A.  A similar analysis was performed by
\citet{dematos19}, who used the VHE $\g$-ray spectrum of Mrk 501 from
HEGRA to attempt to constrain these dust parameters as well.  Our
Model B results are consistent with theirs.  Since the HEGRA spectrum
for Mrk 501 extended to 20 TeV, they were able to a rule out a single
PAH dust component; at least two dust components were required to
explain the $\g\g$ absorption seen in this VHE spectrum.  As
\citet{dematos19} discuss, the upcoming Cherenkov Telescope Array
(CTA) will reach yet higher $\g$-ray energies, and may be able to
provide stronger constraints on the dust parameters.

The temperature we find for the warm, large grains,
$T_1\approx60\ \Kelvin$ is consistent across all models where this was
a free parameter, except Model A.c.  It is quiet a bit higher than the
value used by \citet{finke10_EBL}, which was $T_1= 40\ \Kelvin$.
Model A.c gives a value $T_1=40.5^{+11.7}_{-4.2}\ \Kelvin$, consistent
with \citet{finke10_EBL}.  This model has $H_0$ and $\Omega_m$ as 
free parameters, and returns greater uncertainty for this
parameter.

\subsection{Alternative SFRD and IMF parameterizations}
\label{altSFRDIMF}

We wished to test our model for sensitivity to the IMF and
the SFRD parameterization.  The results of our model with the Salpeter
IMF (Model D) can be seen in Figure \ref{SFR_modelABC}.  The results
for Model D agree with \citet{madau14} at $z\la1$, while it is
increasingly below that model at higher $z$.  Model D agrees
reasonably well with the fiducial Model A (with a Baldry-Glazebrook
IMF) at $z\la 4$, but the Salpeter model gives a higher SFRD,
more consistent with the one form \citet{madau17}.  The difference
seems to be that the dust extinction is a free parameter in our model,
which for \citet{madau14} and \citet{madau17} it was not in their
simpler calculation.  We plan to explore different IMFs in more detail
in a future publication.  The EBL intensity for Model D is still quite
low as with Model A, as is the absorption optical depth in the $0.03 <
z < 0.23$ redshift bin (cf. Section \ref{fiducialcompare}).

We also want to test that the results do not depend strongly on the
SFRD parameterization.  In the right panel of Figure
\ref{SFR_modelABC} we present our results with our standard MD14 SFRD
parameterization and our piecewise SFRD parameterization (Equation
[\ref{piecewiseSFReqn}]).  Our results are again very similar for
$z\la4$, and the piecewise SFRD model at $z\ga4$ is higher than ours.
The model results do have some sensitivity to the SFRD
parameterization as well.  Indeed, their seems to be some degeneracy
between the SFRD and dust parameters.  This is demonstrated in Figure
\ref{dustext_fig}, where clearly the dust extinction for Models A and
E differ significantly.  The greater the dust extinction, the greater
the SFRD must be for there to be more light to make up for that extra
absorption and fit the luminosity density data.  The EBL intensity and
absorption optical depths for Model E are very similar to those for
the fiducial Model A.

\subsection{Cosmological Parameters}
\label{cosmoparams}

Since the Model A $z=0$ EBL intensity is below the lower limits from
galaxy counts at many wavelengths (Figure \ref{EBLintensity}) and
$\g$-ray absorption optical depth is consistently below the observed
data (Figure \ref{tauplot}) we ran model fits with the cosmological
parameters $H_0$ and $\Omega_m$ free in the fit, to see if this could
resolve this discrepancy.  We did model fits with both the BG03 (Model
A.c) and Salpeter IMF (Model D.c).  These models did not resolve the
discrepancy.  

Nevertheless, it is interesting to compare the results of these
  cosmological parameters to values determined with other methods.
For Model A.c, Our value of
$H_0=69.8^{+3.6}_{-3.2}\ \km\ \s^{-1}\ \Mpc^{-1}$ is consistent with
measurements in the near universe from Type Ia supernovae
\citep[$73.2\pm1.3\ \km\ \s^{-1}\ \Mpc^{-1}$;][]{riess22} and in the
far universe from the CMB
\citep[$67.4\pm0.5\ \km\ \s^{-1}\ \Mpc^{-1}$;][]{planck20}.  The
values of $\Omega_m=0.23^{+0.3}_{-0.4}$ is marginally inconsistent
(between $2\sigma$ and $3\sigma$) with the values measured from Type
Ia supernovae \citep[$0.327\pm0.016$;][]{riess22} and the CMB
\citep[$0.315\pm0.007$;][]{planck20}.  For Model D.c, the value of
$H_0=79.4^{+8.1}_{-4.3}\ \km\ \s^{-1}\ \Mpc^{-1}$ is $1.4\sigma$ of
from the Type Ia supernova value, and $2.8\sigma$ from the CMB value;
the value of $\Omega_m=0.29\pm0.04$ is consistent with both the Type
Ia supernova and CMB values.  It is interesting to note that for our
two models with varying cosmological parameters, one is completely
consistent with $H_0$ but slightly discrepant with $\Omega_m$; and
{\em vice versa} for the other.  The main difference between Models
A.c and D.c, with different IMFs, is the number of low-mass stars
produced.  It seems that light produced by low-mass stars in this
model has some sensitivity to cosmological parameters.

\citet{dominguez19} used the same EBL $\g$-ray
absorption data we use (Section \ref{graydatasection}) to measure
cosmological parameters, and their results were
$H_0=67.4^{+6.0}_{-6.2}\ \km\ \s^{-1}\ \Mpc^{-1}$ and
$\Omega_m=0.14^{+0.06}_{-0.07}$.  Our errors are much smaller than
theirs, likely due to (1) their inclusion of systematic uncertainty,
which we do not include; and (2) further constraints impossed on our
EBL model from stellar models and other physical constraints.
Our values from Model A.c are within $1.3\sigma$ of those from
\citet{dominguez19}, while our values for Model D.c are within
$2.1\sigma$ of theirs.  Our values from Model A.c are clearly more
consistent with this previous work, including the low value
for $\Omega_m$.

We emphasize that the cosmological parameters $H_0$ and
$\Omega_m$\ from our model fits should not be considered measurments
of these cosmological parameters.  Unlike the work by
\citet{dominguez19}, we do not include systematic uncertainties.
However, it is worth noting the cosmological parameter results depend
on the chosen IMF, and light produced by low mass stars.  This can
inform future efforts to constrain cosmological parameters with the
EBL.

\section{Discussion}
\label{discussion}

We have presented an update to the EBL model of \citet{finke10_EBL}.
The updated models used detailed PEGASE.2 stellar models, and track
the evolution of the mean gas-phase metallicity and 
  stellar mass density of the Universe.  We have allowed the dust
extinction to evolve with redshift, and used different SFRD
parameterizations and IMFs.  We have fit this model to a wide variety
of data, including luminosity density and $\g$-ray absorption data.
Our modeling has allowed us to measure the SFRD of the Universe with
$\approx 10\%$ uncertainty for the previous $\approx 12.9$\ Gyr, or
$>90$\% of the history of Universe (Section \ref{physimplication}).
The SFRD did show some dependence on the IMF and SFRD parameterization
(Section \ref{altSFRDIMF}).  A more detailed exploration of IMFs will
be presented in a future publication.  We have also strongly
constrained the photon escape fraction from interstellar dust in the
universe and its evolution with redshift, although this also showed
sensitivity to the IMF and SFRD parameterization.  These discrepancies
are mainly at $z\ga3$ (Figures \ref{SFR_modelABC} and
\ref{dustext_fig}).  In Figure \ref{lambda_z_cover_fig} one can see
that at $z\ga1$, the luminosity density data becomes restricted to
$\lambda\la3\ \mu\m$ at $z\ga1$.  This likely accounts for the
uncertainty at high $z$.  The longer wavelength luminosity density
data can nail down the dust emission, which strongly constrains the
dust absorption.  At high $z$ where the longer wavelength data is
lacking, the dust extinction is less constrained, and hence the SFRD
is as well, since there is a degeneracy between the SFRD and dust
extinction.  Luminosity density measurements for $\lambda \ga
3\ \mu\m$ at $z\ga 1$ could help reduce these uncertainties.

Of the data used in the fit, the luminosity density data has by far
the strongest contraining power, due to the smaller uncertainty and
large collection of data (see Table \ref{lumdenstable}). Our fiducial
model (Model A) is a reasonably good fit to almost all of the data
used in the fit.  Some exceptions include the $\approx 1$ --
$3\ \mu\m$ luminosity density at $z>3$ and the EBL absorption optical
depth at $z<0.6$ (Section \ref{fiducialcompare}).  The $z=0$ EBL
intensity predicted by our model is very close to the lower limits
from galaxy counts, and in some cases, is below those limits.  All of
these discrepancies (high-$z$ luminosity density, $z=0$ EBL intensity,
and low-$z$ $\g$-ray absorption) are persistent for different IMF
and SFRD parameterizations.  The EBL intensity and $\g$-ray absorption
discrepancies cannot be resolved by allowing cosmological parameters
$H_0$ and $\Omega_m$ to be free in the fit (Section
\ref{cosmoparams}).

Since the luminosity density data constrain the model so much stronger
than the other data, the discrepancies described above could be
interpreted as a discrepancy between the luminosity density
measurements and the $\g$-ray absorption optical depth measurements.
Since the $\g$-ray absorption is sensitive to {\em all} the light in
the universe, it is possible there is some light not being picked up
in the luminosity density galaxy surveys.  Indeed, there is some
evidence that the galaxy surveys are missing light at large radii from
galaxies in the near-IR from CIBER \citep{cheng21} and in the optical
from the Large Binocular Telescope \citep{trujillo21} and the Hubble
Ultradeep Field \citep{kramer22}.  We will explore the discrepancies
between the $\g$-ray absorption and luminosity densities in a future
publication and attempt to quantify the light missing in galaxy
surveys.

Similar modeling work was done by \citet{andrews18}.  Our EBL mobel
$z=0$ intensity result is significantly below theirs in the stellar
component, where $\lambda \la 6\ \mu\m$.  The primary reason for this
seems to be our inclusion of data with lower luminosity density in
this wavelength range, especially at higher redshifts, as they use the
luminosity density data from \citet{andrews17}.  This is demonstrated
in the $z=0.9$ panel of Figure \ref{LD_modelA_staticz}, where the data
from \citet{tresse07} is below the data from \citet{andrews17}; our
modeling is more consistent with the former.

Observations with the upcoming CTA should result in measurements in
$\g$-ray absorption by the EBL with high precision out to $z\approx2$
with blazars \citep{abdalla21}.  Observations of very nearby $\g$-ray
sources could potentially constrain the EBL out to $100\ \mu\m$
\citep{franceschini19}.  This could provide new insights into the
discrepancies between the EBL inferred from galaxy surveys and from
$\g$-ray absorption observations.

\acknowledgements 

We are grateful to the referee for a constructive report that has
  improved this paper.  J.D.F. was supported by NASA through contract
S-15633Y and through the {\em Fermi} GI program.  A.D. is grateful for
the support of the Ram{\'o}n y Cajal program from the Spanish
MINECO. A.S.L. acknowledges support from Swiss National Science
Foundation.


\bibliographystyle{apj}
\bibliography{grb_ref,references,ULX_ref,gravwave_ref,mypapers_ref,EBL_ref}

\end{document}